\documentclass[graybox]{svmult}

\usepackage{type1cm}        
\usepackage{makeidx}         
\usepackage{graphicx}        
\usepackage{multicol}        
\usepackage[bottom]{footmisc}

\usepackage{newtxtext}       %
\usepackage[varvw]{newtxmath}       

\usepackage{hyperref}
\usepackage{cprotect}

\makeindex             

\begin{document}

\title*{Information flows in nanomachines}

\author{Juan M.R. Parrondo, Jorge Tabanera-Bravo, Federico Fedele, and Natalia Ares}

\institute{Juan M.R. Parrondo
\at Departamento de Estructura de la Materia, F\' isica T\' ermica y Electr\' onica, and GISC.\\ Universidad Complutense de Madrid, 28040-Madrid, Spain,
\email{parrondo@ucm.es}
\and
Jorge Tabanera-Bravo \at Departamento de Estructura de la Materia, F\' isica T\' ermica y Electr\' onica, and GISC.\\  Universidad Complutense de Madrid, 28040-Madrid, Spain,   
\email{jorgetab@ucm.es}
\and
Federico Fedele
\at Department of Engineering Science, University of Oxford, Oxford OX1 3PJ, United Kingdom,
\email{federico.fedele@eng.ox.ac.uk}
\and
Natalia Ares
\at Department of Engineering Science, University of Oxford, Oxford OX1 3PJ, United Kingdom,
\email{natalia.ares@eng.ox.ac.uk}
}

\maketitle

\abstract{Nanomachines are often interpreted as autonomous Maxwell's demons or information engines. This interpretation is appealing from a conceptual viewpoint but sometimes lacks a rigorous and precise formulation. Moreover, it is unclear whether this interpretation is truly useful for analyzing the performance of nanomachines or designing more efficient ones. The concepts of mutual information and information flows provide a quantitative and rigorous account of the role of information in physical systems. In this chapter, we present a pedagogical review of these concepts and apply them to a specific energy transducer at the nanoscale, which couples the tunnel transport of electrons through a carbon nanotube with its mechanical oscillations.}

\section{Introduction}

In 1867, Maxwell introduced his celebrated demon as a Gedanken experiment, in which a `neat-fingered being' capable of observing and operating the microscopic world could defy the second law of thermodynamics and, for instance, extract work from a single thermal bath \cite{Leff}. Single-molecule machines, designed to move and/or perform work at the nanoscale, instantly bring the Maxwell demon to mind.

In Maxwell's original idea, the demon observes the velocity and position of molecules in two gases separated by a wall. Based on these observations, the demon selectively opens and closes a gate in the wall. Specifically, the demon allows fast molecules from the left gas to pass to the right gas and slow molecules from the right gas to pass to the left gas. By doing so, the demon establishes a heat flow from the left to the right and can sustain this flow even against a temperature gradient, without the need of performing work \cite{Leff}.

The crucial aspect of the Maxwell demon is the information gathered during the measurement and subsequently used in the operation of the gate. In this sense, the demon is implementing a feedback control protocol. The first problem that the Maxwell demon poses to thermodynamics is how to incorporate information into the second law or, in other words, how to find a second law for feedback processes \cite{sagawa2012thermodynamics,Parrondo2015Feb, parrondo2023}. The solution was partly given by a stylized version of the demon due to Szil\'ard \cite{Szilard1929}, where an external agent performs a binary measurement, gathering one bit of information, and extracts a work $kT\ln 2$ from a thermal bath at temperature $T$, $k$ being the Boltzmann constant. A more general relationship between the information obtained by the external agent in the measurement and the extracted work (or, equivalently, the entropy reduction) was recently revealed by Sagawa and Ueda \cite{Sagawa2008}, using an interesting concept from information theory: the mutual information between two random variables \cite{Cover}, which we explain in detail in section \ref{sec:info}.

The second problem posed by the Maxwell demon is the restoration of the second law by considering the physical nature of the demon \cite{Parrondo2015Feb} and the heat dissipation or entropy cost associated with its operation. The common wisdom was that there is an unavoidable entropy cost in the measurement \cite{Leff} but, in the 1970s, Charles Bennett applied  Rolf Landauer's ideas on the thermodynamics of information devices \cite{Landauer1961} and introduced an unexpected twist by showing that, in certain setups, the measurement can be carried out with no heat dissipation \cite{Bennett1982b}. In these cases, the restoration of the second law is due to an unavoidable heat dissipation in the resetting of the demon's memory, i.e., in the erasure of the information registered in the measurement. These ideas formed the basis for a new field called {\em thermodynamics of information}, which has benefited from more recent results in non-equilibrium statistical mechanics and stochastic thermodynamics \cite{sagawa2012thermodynamics,Parrondo2015Feb, parrondo2023}. 

The purpose of this chapter is not to provide a review of the thermodynamics of information. The reader can find a more exhaustive introduction to the field in Refs.~\cite{sagawa2012thermodynamics,Parrondo2015Feb,parrondo2023}. Here we focus on the results of the theory that can be applied to autonomous machines. We first introduce in section \ref{sec:info} the main concepts from information theory, namely, Shannon entropy and the mutual information between two random variables. In section \ref{sec:shannon}, we discuss the relationship between Shannon entropy and thermodynamic entropy in stochastic systems described by master equations. There we show that the Shannon entropy of a non-equilibrium system in contact with thermal and particle reservoirs possesses similar properties as the thermodynamic entropy and, in particular, can be used to define a non-equilibrium free energy that assesses the amount of energy that can be converted into useful work. These tools are used in section \ref{sec:thermo} to analyze the thermodynamic properties of different setups where information plays a relevant role: feedback processes and autonomous and driven bipartite systems. Finally, in section \ref{sec:nanotubes} we apply these results to a specific setup based on carbon nanotubes where mechanical oscillations and electron transport are coupled.

\section{Shannon entropy and mutual information}
\label{sec:info}

In 1948, Claude E. Shannon introduced a measure of the uncertainty of a random variable $X$, which could describe, for instance, a message in a communication channel, the microstate of a physical system, or the outcome of a measurement \cite{Cover}. The Shannon uncertainty, also known as Shannon entropy, of a random variable $X$  is defined as
\begin{equation}
    S(X) =S(\rho_X)\equiv -\sum_x \rho_X(x) \log  \rho_X(x)
    \label{eq:entropy_definition}
\end{equation}
where $\rho_X(x)$ is the probability distribution of $X$ and the sum runs over all possible values $x$ that the variable can take on (for continuous random variables, one must replace the sum with an integral, and the entropy is defined up to an additive constant \cite{Cover}). We have used two alternative notations for the Shannon entropy or uncertainty. The first one, $S(X)$, is commonly used in information theory and more appropriate for situations with several random variables (see below). The second one, $S(\rho_X)$ expresses explicitly the dependence of Shannon entropy on the probability distribution $\rho_X$ and is more usual in statistical physics. Shannon uncertainty is measured in \textit{bits} if the logarithm in Eq.~\eqref{eq:entropy_definition} is in base $2$, and in \textit{nats} if the natural logarithm is used. Moreover, if we multiply \eqref{eq:entropy_definition} by the Boltzmann constant, Shannon entropy is then measured in units of physical entropy (energy divided by absolute temperature). In this work, we will always express Shannon entropy or Shannon uncertainty in these units or, more precisely, in nats times the Boltzmann constant $k$.

Information is defined as the reduction in the uncertainty of a random variable $X$ when we inquire about it by, for instance,  performing a measurement or asking a question to somebody who knows $X$. The outcome of the measurement or the answer to the question is a random variable $Y$, not necessarily equal to $X$. For example, if somebody knows the result $X=1,2,\dots,6$ of a roll of a die and we ask her whether $X$ is odd, then the answer $Y$ takes on only two possible values, `yes' or `no'. The information that the answer $Y$ provides about the random variable $X$  is the reduction of its uncertainty, i.e.,
\begin{equation}\label{eq:def_mutual}
    I(X;Y) = S(X) - S(X|Y),
\end{equation}
and is called the \textit{mutual information} between $X$ and $Y$. Here $S(X|Y)$ is the uncertainty of the posterior distribution $\rho_{X|Y}(x|y)$ averaged over the measurement outcomes $y$:
\begin{equation}
    S(X|Y) = \sum_{y} \rho_Y(y) \left[-\sum_{x} \rho_{X|Y}(x|y) \log  \rho_{X|Y}(x|y)\right].
\end{equation}
Using the relation $\rho_{XY}(x,y) = \rho_{X|Y}(x|y)\rho_Y(y)$ one obtains
\begin{equation}
    I(X;Y) = \sum_{x,y} \rho_{XY}(x,y)\log \frac{\rho_{XY}(x,y)}{\rho_X(x)\rho_Y(y)}.
\end{equation}
We observe that mutual information is symmetric, that is, the information that $X$ provides about $Y$ equals the information that $Y$ provides about $X$. In addition, mutual information is a measure of the correlation between the two variables: $I(X;Y)\geq 0$ and vanishes when $X$ and $Y$ are uncorrelated. Another way of writing mutual information is
\begin{equation}
    I(X;Y) = S(X) + S(Y) - S(X,Y) \geq 0 \label{eq:information_proerty}
\end{equation}
which shows that entropy is sub-additive: $S(X,Y)=S(X)+S(Y)-I(X;Y)\leq S(X)+S(Y)$. The equality is met if and only if the two random variables $X$ and $Y$ are statistically independent.

If $Y$ is a deterministic function of $X$, which is the case of an error-free measurement, then $S(Y|X)=0$ and $I(X;Y)=S(Y)-S(Y|X)=S(Y)$, i.e., the mutual information is the Shannon entropy of the measurement outcome.

Equations \eqref{eq:def_mutual} and \eqref{eq:information_proerty} express, respectively,  two properties of mutual information that are relevant in thermodynamics. The first property \eqref{eq:def_mutual} states that mutual information is equal to the decrease in entropy resulting from a measurement. The second property \eqref{eq:information_proerty} is useful for connecting the entropy $S(X,Y)$ of a bipartite system described by the random variables $(X,Y)$ with the entropies of its subsystems, $S(X)$ and $S(Y)$. However, in order to understand the physical implications of measurements and other processes involving information, it is still necessary to clarify the relationship between Shannon entropy and thermodynamic entropy.

\section{Shannon and thermodynamic entropies}
\label{sec:shannon}

Shannon entropy coincides with thermodynamic entropy for equilibrium states. Nevertheless, the identification of both magnitudes for systems far from equilibrium is not always correct. For instance, the Shannon entropy of a probability distribution over the microstates of a system is invariant under Hamiltonian evolution and, consequently, does not reproduce the increase in thermodynamic entropy dictated by the second law.
However, in the case of non-equilibrium systems in contact with thermal baths and particle reservoirs, Shannon entropy has the same properties as thermodynamic entropy. This can be proved in the context of Hamiltonian evolution \cite{parrondo2023}, as well as stochastic evolution given by Langevin and/or master equations \cite{Esposito2010_three1,Esposito2010_three2}. We now discuss this important result for a discrete system ruled by a master equation, following Ref.~\cite{Esposito2010_three1}, and later on we extend the discussion to a bipartite hybrid system (continuous and discrete) describing the coupling between the electron transport through a carbon nanotube and its mechanical oscillations.

\subsection{Entropy production in discrete systems}
\label{sec:entropymaster}

Consider a system with discrete states $i$ in contact with several thermostats and chemostats (particle reservoirs) at different temperatures and chemical potentials. The stochastic evolution of the system is described by a master equation:
\begin{equation}\label{master}
    \dot \rho(i,t)=\sum_j \left[ \rho(j,t)\Gamma_{j\to i}-\rho(i,t)\Gamma_{i\to j}\right]
\end{equation}
where $\rho(i,t)$ is the probability that the system is in state $i$ at time $t$ and $\Gamma_{i\to j}$ is the probability transition rate from state $i$ to state $j$ (more precisely, the probability to observe the transition $i\to j$ in a short time interval of duration $\Delta t$ is $\rho(i,t)\Gamma_{i\to j}\Delta t$). The master equation can also be written as
\begin{equation}\label{masterj}
    \dot \rho(i,t)=\sum_j  J_{j\to i}(t)
\end{equation}
where
\begin{equation}
    J_{j\to i}(t)\equiv \rho(j,t)\Gamma_{j\to i}-\rho(i,t)\Gamma_{i\to j}
\end{equation}
is the net probability current from state $j$ to state $i$. 
The change in the Shannon entropy  of the system can be written as
\begin{align}
    \dot S(t)&=-\frac{d}{dt}k\sum_i \rho(i,t)\ln \rho(i,t)
    =-k\sum_i\frac{d\,\rho(i,t)}{dt} \ln \rho(i,t)\nonumber \\
    &=-k\sum_{i,j}J_{j\to i} (t)\ln \rho(i,t)
    =k\sum_{i>j}J_{i\to j} (t)\ln\frac{\rho(i,t)}{\rho(j,t)}\label{sxmarkov}
\end{align}
where we have used the normalization of the probability distribution: $\sum_i d\rho(i,t)/dt=0$ in the first step and the anti-symmetry of the current: $J_{i\to j}(t)=-J_{j\to i}(t)$ in the last step. The expression indicates that each transition, $i \to j$, contributes to the increase in Shannon entropy as $k\ln [\rho(i,t)/\rho(j,t)]$.

To relate Shannon entropy with thermodynamic entropy, we need to specify the physical nature of the system and its environment. For simplicity, we will assume that the states of the system do not have internal entropy and their free energy equals their energy $E_i$. The system exchanges energy with several thermal baths at different temperatures and some transitions can be mediated by particle reservoirs with different chemical potentials. In this case, the transition rates obey local detailed balance \footnote{In some setups, a transition can follow different pathways involving different reservoirs. In those situations, local detailed balance is fulfilled by the rates of each pathway \cite{esposito2007}. This is the case of a quantum dot in contact with two reservoirs, as we discuss in section \ref{sec:dot_dynamics}.}:
\begin{equation}\label{db}
    \frac{\Gamma_{i\to j}}{\Gamma_{j\to i}}=e^{-\beta_{ji}\left( E_j-E_i+\Delta F^{{\rm (env)}}_{i\to j}\right)}
\end{equation}
where $\beta_{ji}=\beta_{ij}=1/(kT_{ij})$ is the inverse temperature of the bath that induces the transitions between $i$ and $j$ and $\Delta F^{{\rm (env)}}_{i\to j}$ is the change in the free energy of the environment due to the transition from $i$ to $j$. Since $F=E-TS$ in each reservoir and $E_i-E_j=\Delta E^{{\rm (env)}}_{i\to j}$ is the energy dissipated to the reservoir at  temperature $T_{ij}$, the total change in the  thermodynamic entropy of the environment due to  a transition from $i$ to $j$ is given by
\begin{equation}\label{fenv}
     \Delta  S^{\rm (env)}_{i\to j}=\frac{E_i-E_j-\Delta F^{{\rm (env)}}_{i \to j}}{T_{ij}}
  =k\ln \frac{\Gamma_{i\to j}}{\Gamma_{j\to i}}.
\end{equation}
Then, the entropy of the environment changes at a rate
\begin{equation}
    \dot S_{\rm env}(t)=\sum_{i>j}J_{i\to j}(t) \Delta  S^{\rm (env)}_{i\to j} =
    k\sum_{i>j}J_{i\to j}(t)\ln \frac{\Gamma_{i\to j}}{\Gamma_{j\to i}}.\label{senvapp} 
\end{equation}

If we sum up the change in the thermodynamic entropy of the environment \eqref{senvapp} and in the Shannon entropy of the system \eqref{sxmarkov}, we get a total entropy production that reads
\begin{equation}
     \dot S_{\rm prod}(t)\equiv \dot S_{\rm env}(t)+\dot S(t)= k\sum_{i>j}J_{i\to j} (t)\ln\frac{\rho(i,t)\Gamma_{i\to j}}{\rho(j,t)\Gamma_{j\to i}}\geq 0.
    \label{sprodmarkov}   
\end{equation}
Each term in the sum is positive because the logarithm and the current have the same sign. Hence, $\dot S_{\rm prod}(t)\geq 0$, and the equality is reached only in the case of global detailed balance, where all currents vanish. In summary, we have proved that the Shannon entropy can be added to the environment entropy to yield a positive entropy production. Notice that the environment entropy is well-defined if the environment consists of equilibrium reservoirs. In this case, the change in entropy $\Delta S$ is given by the usual expressions in equilibrium thermodynamics: $\Delta S=\Delta E/T+\mu \Delta N/T$, where $\Delta E$ and $\Delta N$ are the changes in the energy and number of particles of the reservoir, respectively, $T$ is its temperature, and $\mu$ its chemical potential. Hence, for systems in contact with equilibrium reservoirs, Shannon entropy possesses the same properties as thermodynamic entropy even for non-equilibrium states and processes.

\subsection{Energetics and non-equilibrium free energy}

All the previous equations hold for time-dependent transition rates $\Gamma_{i\to j}(t)$. In particular, they are valid for driven physical systems, with energies $E_i(\lambda)$ depending on a parameter $\lambda$ that an external agent modifies according to a given protocol $\lambda(t)$ with $t\in [0,\tau]$, $\tau$ being the duration of the protocol.
In this case, we can split the change in the average energy of the system $E(t)\equiv \sum_i p(i,t)E_i(\lambda(t))$ into two terms:
\begin{equation}\label{edot}
    \dot E(t)=\sum_i \left[ \frac{ dp(i,t)}{dt}E_i(\lambda(t))+
  p(i,t)\frac{dE_i(\lambda(t))}{dt}  \right].
\end{equation}
The first term is the exchange of energy with the environment. It can be expressed as a sum of the energy transfer in each transition, following an analogous argument as the one used in Eq.~\eqref{sxmarkov}:
\begin{equation}
    \sum_i  \frac{ dp(i,t)}{dt}E_i(\lambda(t))=\sum_{i>j}J_{i\to j}(t)\left[E_j(\lambda(t))-E_i(\lambda(t))\right].
    \end{equation}
This quantity can be further split into two terms  using \eqref{fenv}:
\begin{align}
    \sum_i  \frac{ dp(i,t)}{dt}E_i(\lambda(t))&=-\sum_{i>j}J_{i\to j}(t)\left[T_{ij}\Delta S_{i\to j}^{\rm (env)}+\Delta F_{i\to j}^{\rm (env)}\right]\nonumber \\ &=\dot Q(t)+\dot W_{\rm chem}(t).
    \end{align}
Here, the heat $\dot Q (t)\equiv -\sum_{i>j}J_{i\to j}(t)T_{ij}\Delta S_{i\to j}^{\rm (env)}$ is the energy that changes the entropy of the environment, whereas the rest of the energy is usually called chemical work,  $\dot W_{\rm chem}(t)\equiv -\sum_{i>j}J_{i\to j}(t)\Delta F_{i\to j}^{\rm (env)}$. This work is energy that the reservoirs transfer to the system without affecting their entropy. Chemical work is relevant when some transitions are mediated by chemostats, as occurs in chemical motors \cite{grelier2023}, as well as when the system and the reservoirs exchange particles. This is the case of the setup that we discuss in section \ref{sec:nanotubes}, where a quantum dot exchanges electrons with two electrodes.

Finally, the last term in \eqref{edot} is the energy that the system takes from the external agent, and is usually called mechanical work or simply work:
\begin{equation}\label{edot2}
   \dot W(t)\equiv\sum_i 
  p(i,t)\frac{dE_i(\lambda(t))}{dt}.
\end{equation}
These energy exchanges can be combined into a single equation, which is the mathematical expression of the first law of thermodynamics\footnote{Note that the overdot on heat and work does not indicate the time derivative of a given quantity, but a rate of energy exchange.}:
\begin{equation}\label{firstlaw}
    \dot E(t)=\dot Q(t)+\dot W(t)+\dot W_{\rm chem}(t).
\end{equation}

For an isothermal process (i.e., when all the baths are at the same temperature $T$), the heat is $\dot Q(t)=-T\dot S_{\rm env}(t)$. In this case,
from the first law \eqref{firstlaw} and the second law \eqref{sprodmarkov}, one can immediately obtain
\begin{equation}\label{isotherm}
T\dot S_{\rm prod}(t)=-\dot Q(t)+T\dot S(t)=\dot W(t)+\dot W_{\rm chem}(t)-\dot {\cal F}(t)\geq 0
\end{equation}
where
\begin{equation}
    {\cal F}(t)\equiv E(t)-TS(t)
\end{equation}
is called {\em non-equilibrium free energy}. It can be defined in general for any physical system with Hamiltonian ${\cal H}(x)$  and probabilistic state $\rho_X(x)$ as  the average energy minus the Shannon entropy multiplied by the temperature \cite{Parrondo2015Feb}:
\begin{equation}\label{noneqf}
    {\cal F}(\rho_X,{\cal H})\equiv \langle{\cal H}(X)\rangle_{\rho_X} - TS(\rho_X).
\end{equation}
We will also use the notation ${\cal F}(X)$ for the non-equilibrium free energy of a system whose microstate is the random variable $X$ with probability distribution $\rho_X(x)$.
This definition is valid for probability distributions $\rho(i)$ defined over micro- or meso-states $i$ with energy $E_i$, as well as for quantum density matrices. Like in the case of isothermal processes connecting equilibrium states, the non-equilibrium free energy tells us the amount of work that can be extracted from non-equilibrium states. Indeed, integrating Eq.~\eqref{isotherm} over the whole process $t\in [0,\tau]$, we obtain
\begin{equation}\label{wf}
    W_{\rm tot}\geq \Delta {\cal F}
\end{equation}
where $W_{\rm tot}=W+W_{\rm chem}$ is the total work performed by the external agent and the reservoirs on the system, and  
\begin{equation}\label{df}
   \Delta {\cal F}={\cal F}\left[\rho_X(x;\tau),{\cal H}(x;\tau)\right]-{\cal F}\left[\rho(x;0),{\cal H}(x;0)\right]
\end{equation}
is the increase in non-equilibrium free energy. Here $\rho_X(x;t)$ and ${\cal H}(x;t)$ are, respectively, the probabilistic state of the system and the Hamiltonian at time $t$. Thus, we can extract work, $W_{\rm tot}<0$, by driving a system from a non-equilibrium state with high free energy to a state with lower non-equilibrium free energy. See \cite{Parrondo2015Feb} for a specific protocol that is able to extract the maximum amount of work $-\Delta {\cal F}$.
These results are also valid for continuous systems ruled by Langevin-like equations, as well as for a wide family of Hamiltonian systems  \cite{parrondo2023}.

\section{Thermodynamics of information}
\label{sec:thermo}

We have proven that the change in Shannon entropy of a system can be added to the change in thermodynamic entropy of its environment, resulting in a total entropy production that is positive and only vanishes for reversible processes, as shown by Eq.~\eqref{sprodmarkov}. Shannon entropy is also used to define non-equilibrium free energy in Eq.~\eqref{noneqf}, which establishes a lower bound on the work required to complete an isothermal process connecting non-equilibrium states, as shown by Eq.~\eqref{wf}. 

In this section, we analyze the interplay between information, entropy,  and energy in different physical situations, utilizing the properties of the Shannon entropy introduced in section \ref{sec:info}. Processes in which information plays a significant role always involve non-equilibrium states, such as those resulting from a measurement. In this sense, thermodynamics of information can be regarded as a branch of non-equilibrium thermodynamics that deals with a specific class of non-equilibrium states.

We begin by discussing the effect of feedback on the second law of thermodynamics, which is present in the original Maxwell demon and the Szil\'ard engine. Next, we incorporate the physical nature of the demon to restore the original second law. For this purpose, we study bipartite systems, constituted by the machine and the demon. We first consider bipartite systems driven by an external agent who switches on and off the interactions between the machine and the demon and restores the state of the demon. This external agent drives the system according to a non-feedback protocol, that is, they cannot measure the bipartite system nor use any information of its actual state. In this way, the bipartite system undergoes a normal non-feedback thermodynamic process and the second law is restored. Nevertheless, the process can be interpreted as the demon gathering information from the machine and operating according to that information. Finally, we consider the case of bipartite autonomous systems in a non-equilibrium steady state (NESS). In this case, which is probably the most relevant for nanomachines, we can still interpret one of the two systems as gathering information from the other and implementing some kind of feedback, using a new concept: information flows. 

\subsection{Isothermal feedback processes}

The simplest setup in which information affects the limitations imposed by thermodynamics is a process where an external agent performs a measurement on a system and implements a protocol based on the measurement outcome $Y$. According to \eqref{eq:def_mutual}, the measurement reduces the Shannon entropy of the system by an amount $I(X;Y)$, where $X$ is the microstate of the system. If the measurement does not affect the energy of the system, this reduction in entropy  leads to an increase in the non-equilibrium free energy:
\begin{equation}
    \Delta {\cal F}_{\rm meas} \equiv  {\cal F}_{\rm post}-{\cal F}_{\rm pre}= TI(X;Y).
\end{equation} 
where ${\cal F}_{\rm post}$ and ${\cal F}_{\rm pre}$ are, respectively, the non-equilibrium free energies immediately before and after the measurement.

\begin{figure}[t]
    \centering
    \includegraphics[width=0.7\linewidth ]{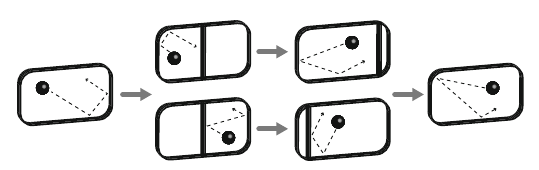}
    \caption{Sketch of the Szil\'ard engine:  a piston is inserted in the middle of a box containing a single particle gas. By measuring the half where the particle lies after the insertion, one can perform a reversible isothermal expansion and extract an energy $kT\ln 2$ from the thermal bath.}
    \label{fig:szilard}
\end{figure}

In principle, this increase may require work and heat dissipation, as we will discuss in the next subsection. Here, we are solely interested in finding how the second law \eqref{wf} should be modified to incorporate the information acquired in the measurement. The modification simply consists in adding the increment in free energy $\Delta {\cal F}_{\rm meas}$ to the available free energy that can be converted into work. 
More precisely, the work before the measurement obeys $W_{\rm pre}\geq {\cal F}_{\rm pre}-{\cal F}_0$, where ${\cal F}_0$ is the initial free energy, and the work after the measurement is bound as $W_{\rm post}\geq {\cal F}_{\tau}-{\cal F}_{\rm post}$, where  ${\cal F}_{\tau}$ is the free energy at the end of the process. 
If we do not take into account the work performed in the measurement, the  non-equilibrium second law \eqref{wf}, that is, the total work necessary to complete the process is bound as \cite{parrondo2023}
\begin{equation}\label{second_law_feedback}
     W_{\rm fb}=W_{\rm pre}+W_{\rm post }\geq   \Delta {\cal F}-TI(X;Y).
\end{equation}
For a cyclic process, $\Delta {\cal F}=0$ but, if the measurement provides some information about the microstate of the system, one can extract work ($W_{\rm fb}<0$), in apparent contradiction with the second law of thermodynamics. A simple example is the aforementioned Szil\'ard engine 
\cite{Szilard1929}, in which a particle is in a container of volume $V$ whose walls are at temperature $T$. A piston is inserted in the middle of the container and the demon measures in which half the particle lies, i.e., the outcome $Y$ is either left or right with equal probability. If the measurement 
has no errors, then $I(X;Y)=S(Y)=1\,\mbox{ bit}=k\ln 2$.  This information can be used to implement a reversible isothermal expansion extracting an energy $kT\ln 2$ from the thermal bath, in agreement with the second law \eqref{second_law_feedback}. The process is sketched in figure  \ref{fig:szilard}.

An interesting problem is whether the bound in \eqref{second_law_feedback} can be reached. This is the case of the Szil\'ard engine, for instance. Protocols that saturate the bound are called optimal feedback protocols or optimal Maxwell demons  \cite{Horowitz2011b}. In  \cite{Horowitz2011}, we show that they must be operationally reversible, i.e., the system must go through the same probabilistic states when the protocol is run forward and backward in time. This implies that optimal protocols are quasistatic, as in standard thermodynamics, but must also reproduce post-measurement probabilistic states,  a requirement that can be difficult to fulfill and yields highly non-trivial protocols \cite{Horowitz2011b}.

\subsection{Driven bipartite systems}

We now consider the physical nature of the demon. As a physical system, the demon is described by a microscopic state $z$ comprising the positions and momenta of all its molecules. Its evolution and interaction with the system are ruled by a Hamiltonian 
${\cal H}_{\rm demon}(z)+{\cal H}_{\rm int}(x,z;t)$, where $x$ denotes the microstate of the controlled system. Still, an external agent is needed to switch on and off the interaction between the demon and the system in the measurement and in the feedback operation. This external agent, however, carries out a protocol on the bipartite system (system$+$demon)  that does not use any information about the state of the system. Hence, the usual second law \eqref{wf} holds. 
According to equation \eqref{eq:information_proerty}, the entropy of the bipartite system can be written as
\begin{equation}\label{si}
    S(X(t),Z(t))=S(X(t))+S(Z(t))-I(X(t);Z(t)),
\end{equation}
and the non-equilibrium free energy is
\begin{equation}\label{fi}
   {\cal F}(X(t),Z(t))= {\cal F}(X(t))+{\cal F}(Z(t))+\langle {\cal H}_{\rm int}(X,Z;t)\rangle +TI(X(t);Z(t)).
\end{equation}

\begin{figure}[h!]
    \centering
    \includegraphics[width=.3\linewidth ]{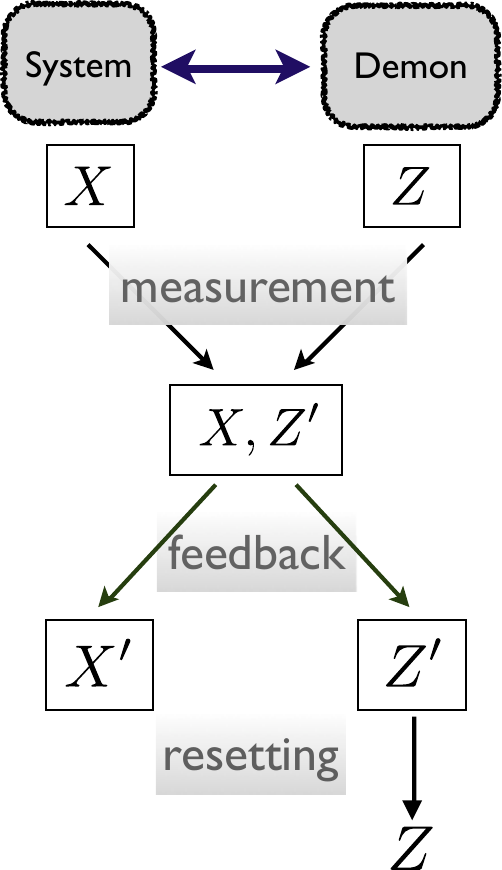}
    \caption{Evolution of the state of the system and the demon in the different stages of a feedback process.}
    \label{fig:measurement}
\end{figure}

Consider now the feedback isothermal process depicted in figure \ref{fig:measurement}. We assume that the demon and the system start in random states $X$ and $Z$, respectively, without any interaction ${\cal H}_{\rm int}=0$ and correlation, $I(X;Z)=0$.
The measurement is an interaction between the demon and the system that, ideally, does not alter the state $X$ of the system. The state of the demon does change to $Z'$ and correlates with the state of the system generating a mutual information $I(X;Z')>0$. The outcome of the measurement $Y$ is a function of the microstate of the demon $Z'$:  $Y=y(Z')$. If all the information about $X$ is provided by the measurement outcome $Y$, then  the conditional probabilities obey $\rho_{X|Z'}(x|z')=\rho_{X|Y}(x|y(z'))$, and it is not hard to prove that the mutual information verifies
\begin{equation}
  I(X;Z')=  I(X;Y).
\end{equation}
If the interaction is switched off after the measurement, then the change in non-equilibrium free energy due to the measurement is $\Delta {\cal F}_{\rm meas}={\cal F}(Z')+TI(X;Z')-{\cal F}(Z)$. 
This change requires a  work bound by $\Delta{\cal F}_{\rm meas}$:
\begin{equation}\label{wmeas}
    W_{\rm meas}\geq \Delta{\cal F}_{\rm meas}=\Delta{\cal F}_{\rm demon}+TI(X;Y)
\end{equation}
where $\Delta{\cal F}_{\rm demon}={\cal F}(Z')-{\cal F}(Z)$ is the change in the non-equilibrium free energy of the demon due to the measurement. Notice that the measurement can be carried out without any work and heat dissipation if the free energy of the demon decreases and compensates for the term due to the mutual information \cite{Bennett1982b}. 

After the measurement, the demon implements the feedback process, where the system is driven from state $X$ to state $X'$. We assume that the demon state does not change during the feedback protocol. If all correlations are
exhausted, meaning that $I(X';Z')=0$,  and the interaction is switched off at the end of the process, then the change in non-equilibrium free energy in the feedback process is
$\Delta {\cal F}_{\rm fb}=\Delta{\cal F}_{\rm system}-TI(X;Z')$,
where $\Delta{\cal F}_{\rm system}={\cal F}(X')-{\cal F}(X)$, and the work is bound as:
\begin{equation}\label{wfb}
    W_{\rm fb}\geq \Delta{\cal F}_{\rm fb}=\Delta{\cal F}_{\rm system}-TI(X;Y).
\end{equation}
This inequality coincides with \eqref{second_law_feedback}, as expected. However, we have obtained a new result: to achieve the equality in both equations, \eqref{second_law_feedback} and \eqref{wfb}, it is necessary to exploit all the correlations $I(X;Z')=I(X;Y)$ generated in the measurement. Notice also that, in the derivation of \eqref{second_law_feedback},  mutual information appeared as the reduction of uncertainty in the system due to the measurement, whereas here it is interpreted as a measure of the correlation between the demon and the system.

For a cycle,  $\Delta{\cal F}_{\rm system}=0$, and the amount of work that can be extracted from the thermal bath is $TI(X;Y)$. If we perform a measurement without work, i.e., decreasing the free energy of the demon by $\Delta {\cal F}_{\rm demon}=-TI(X;Y)$, then we have an information engine that can extract work from a single thermal bath. However, there is no contradiction with the second law, since the final state of the demon $Z'$ is not equal to the initial state $Z$. In other words, the process is not cyclic unless we add a third step where the memory of the demon is reset. The resetting changes the free energy as $-\Delta {\cal F}_{\rm demon}$ and needs a work that verifies
\begin{equation}\label{wreset}
   W_{\rm reset}\geq -\Delta {\cal F}_{\rm demon}.
\end{equation}
Summing up the three bounds, \eqref{wmeas},\eqref{wfb}, and \eqref{wreset}, we conclude that the total work $W_{\rm tot}=W_{\rm meas}+W_{\rm fb}+W_{\rm reset}$ is always positive in a cycle ($X'=X$), as dictated by the second law of thermodynamics.

In light of this discussion, we conclude that the Maxwell demon consists of the creation and subsequent exploitation of correlations between two systems: one acting as a demon and the other as a ``working substance''. This interpretation applies to non-autonomous nanomachines described by bipartite systems: the intervals of time where the mutual information between the two systems, $I(X(t);Y(t))$,  increases can be identified as measurements, whereas those intervals where the mutual information decreases can be interpreted as feedback processes where the information is converted into work \cite{Horowitz2014Jul}. This analysis can reveal the efficiency of each stage, measurement, feedback, and resetting, as well as whether the correlations generated by the measurement are optimally exploited in the feedback stage of the process. 

Another interesting aspect of the previous analysis is that we can leave out the last stage of the cycle ---resetting the demon's memory--- if the memory has initially a high non-equilibrium free energy. In this case, we can run the process many times extracting work and degrading the memory.   In other words, an ordered memory can be used as fuel. These information reservoirs have been studied in general by Barato and Seifert in   \cite{Barato2014b}. Mandal and Jarzynski 
devised a  specific kinetic model that can extract work from a single thermal bath using as fuel an ordered memory on a tape \cite{Mandal2012}.

If we want to extend the previous analysis to bipartite systems in contact with several reservoirs at different temperatures, we must use entropy instead of free energy. The combination of the second law \eqref{sprodmarkov} and the expression for the Shannon entropy of a bipartite system $(X,Y)$ in terms of the mutual information \eqref{si} yields
\begin{equation}\label{dots}
\dot S_{\rm prod}=\dot S(X)+\dot S(Y)- \dot I(X;Y)+\dot S_{\rm env}\geq 0,
\end{equation}
where now the change in the entropy of the environment depends on the physical properties of the reservoirs.


\subsection{Autonomous bipartite systems: information flow}
\label{sec:bipartite1}

Consider an autonomous bipartite system $(X,Y)$ in a non-equilibrium steady state (NESS). Eqs \eqref{si}, \eqref{fi}, and \eqref{dots} are still valid but useless since all the terms are constant in the first two equations and vanish in the latter.  Information flows try to overcome this problem by focusing on the changes in mutual information due to the evolution of each part of the bipartite system. They are defined as \cite{Horowitz2014Jul,Allahverdyan2009Sep,hartich2014,ehrich2023}
\begin{equation}\label{infoflows}
    \dot I^{X}(t)\equiv \left.\frac{d}{dt'}\right|_{t'=t}I(X(t');Y(t));\quad
\dot I^{Y}(t)\equiv \left.\frac{d}{dt'}\right|_{t'=t}I(X(t);Y(t')).
\end{equation}

Until now, we have used the term  `bipartite system' to refer to any system with two parts described by microstates $(X,Z)$ or $(X,Y)$\footnote{In this subsection, and in the rest of the chapter, we no longer distinguish between the state of the demon and the measurement outcome.}. To establish a connection between the information flows  \eqref{infoflows} and the change in mutual information, we need to impose further restrictions on this definition.
In a bipartite discrete system, transitions where the state of the two subsystems changes simultaneously are not allowed, i.e.,
\begin{equation}
    \Gamma_{(x,y)\to (x',y')}=0 \quad \mbox{if $x\neq x'$ and $y\neq y'$}.
\end{equation}
Hence, one can consider separately the transitions where each of the subsystems changes:
\begin{equation}
    \Gamma^y_{x\to x'}\equiv \Gamma_{(x,y)\to (x',y)}
\end{equation}
and
\begin{equation}
    \Gamma^x_{y\to y'}\equiv \Gamma_{(x,y)\to (x,y')}.
\end{equation}
The master equation \eqref{masterj} can be split into two sums corresponding  to the evolution of systems $X$ and $Y$, respectively,
\begin{equation}
    \dot \rho(x,y;t)=\sum_{x',y'} \left[ J^y_{x'\to x}(t)+ J^x_{y'\to y}(t)\right]
\end{equation}
where the current due to changes in $X$ is
\begin{equation}
     J^y_{x'\to x}(t)\equiv \Gamma^{y}_{x'\to x}\rho(x',y;t)
    -\Gamma^{y}_{x\to x'}\rho(x,y;t)
\end{equation}
and the current corresponding to system $Y$ reads
\begin{equation}
     J^x_{y'\to y}(t)\equiv \Gamma^{x}_{y'\to y}\rho(x,y';t)
    -\Gamma^{x}_{y\to y'}\rho(x,y;t).
\end{equation}

Now every term in Eq.~\eqref{dots} can be decomposed into two parts accounting for the contributions of transitions in systems $X$ and $Y$, respectively. 
For transitions in system $X$, this contribution is
\begin{equation}\label{sprodmark}
  \dot S^X_{\rm prod}{(t)}\equiv  \dot S(X(t))-\dot I^X(t)+\dot S^{X}_{\rm env}(t).
\end{equation}
Each derivative here is expressed in terms of $J^y_{x'\to x}(t)$. Using similar arguments as in the derivation of Eq.~\eqref{sxmarkov}, we  express the change in Shannon entropy as
\begin{equation}\label{sxmark}
  \dot S(X(t))=k\sum_{y}\sum_{x>x'}J^y_{x'\to x}(t)\ln\frac{\rho(x',t)}{\rho(x,t)}.
\end{equation}
It is not difficult to prove that the information flow  defined in \eqref{infoflows} can be written as
\begin{align}
    \dot I^X(t) &= k\sum_{x,y}\sum_{x'}J^y_{x'\to x}(t)\ln\frac{\rho(x,y;t)}{\rho(x,t)\rho(y,t)}\nonumber\\
    &= k\sum_{y}\sum_{x>x'}J^y_{x'\to x}(t)\ln\frac{\rho(x,y;t)\rho(x',t)}{\rho(x',y;t)\rho(x,t)},\label{ixmar}
\end{align}
where we have used the anti-symmetry of the current $J^y_{x'\to x}=-J^y_{x\to x'}$. Finally, 
the entropy increase in the environment due to transitions in $X$ is given by
\begin{equation}\label{senvxmark}
    \dot S^{X}_{\rm env}(t)=k\sum_y\sum_{x>x'}J^y_{x'\to x}(t)\ln \frac{\Gamma^{y}_{x'\to x}}{\Gamma^{y}_{x\to x'}}.
\end{equation}
This expression is a consequence of local detailed balance, as in \eqref{senvapp}. Introducing Eqs.~\eqref{sxmark}, \eqref{ixmar}, and \eqref{senvxmark} into \eqref{sprodmark}, we obtain
\begin{equation}\label{sprodX}
     \dot S^X_{\rm prod}{(t)}=k\sum_y\sum_{x>x'}J^y_{x'\to x}(t)\ln \frac{\rho(x',y;t)\Gamma^{y}_{x'\to x}}{\rho(x,y;t)\Gamma^{y}_{x\to x'}}\geq 0.
\end{equation}
As in Eq.~\eqref{sprodmarkov}, the inequality holds because every term in the sum is positive, given that the current and the logarithm have the same sign.

In the stationary state, the marginal Shannon entropies are constant:  $\dot S(X)=\dot S(Y)=0$, as well as the mutual information: $\dot I(X;Y)=\dot I^X+\dot I^Y=0$. Hence, there is a single quantity characterizing the information flow: $\dot I^Y=-\dot I^X$. The sign of the information flow indicates the role played by each subsystem. In the previous section, we showed that mutual information increases during the measurement and decreases during feedback. Additionally, we observed that, in an ideal measurement, the state of the demon changes and becomes correlated with the state of the system, without altering the latter. Conversely, in the feedback process,  the system changes while, ideally, the state of the demon remains constant. We can apply these observations to the autonomous bipartite system. For instance, if $\dot I^Y$ is positive and $\dot I^X$ is negative, then the evolution of system $Y$, keeping $X$ constant,  leads to an increase in mutual information, while the evolution of system $X$ decreases mutual information. By comparing this scenario with the driven case, we can identify system $Y$ as the demon and system $X$ as the one on which the demon measures and operates.

The inequality \eqref{sprodX} together with the definition of the entropy production \eqref{sprodmark} due to the evolution of $X$, and the analogous equation for system $Y$, yield the following local second laws for the bipartite system in a NESS \cite{Horowitz2014Jul,hartich2014}:
\begin{equation}\label{secondflows}
    \begin{split}
        \dot S^X_{\rm prod}&= \dot S^{X}_{\rm env}+\dot I^Y \geq 0\\
     \dot S^Y_{\rm prod}&= \dot S^{Y}_{\rm env}-\dot I^Y\geq 0.
    \end{split}
\end{equation}
This is the main result of this section. The sum of the two inequalities yields the usual second law for a NESS, namely, that the entropy in the environment increases, $\dot S_{\rm env}\geq 0$. The two inequalities in \eqref{secondflows} impose an extra limitation on the functioning of the system and reveal how the bipartite system operates as an information machine \cite{Horowitz2014Jul,hartich2014}.
If $\dot I^Y$ is positive, $Y$ acts as a demon that measures on $X$. For this purpose, $Y$ uses the entropic resources of its environment, $\dot S^{Y}_{\rm env}\geq \dot I^Y\geq 0 $. As a consequence, system $X$ can reduce the entropy of its reservoirs: $\dot S^{X}_{\rm env}\geq -\dot I^Y$.    Therefore,
subsystem $X$ could apparently defy the second law of thermodynamics due to the information flow from subsystem $Y$.

\begin{figure}
    \centering
    \includegraphics[width=.56\linewidth ]{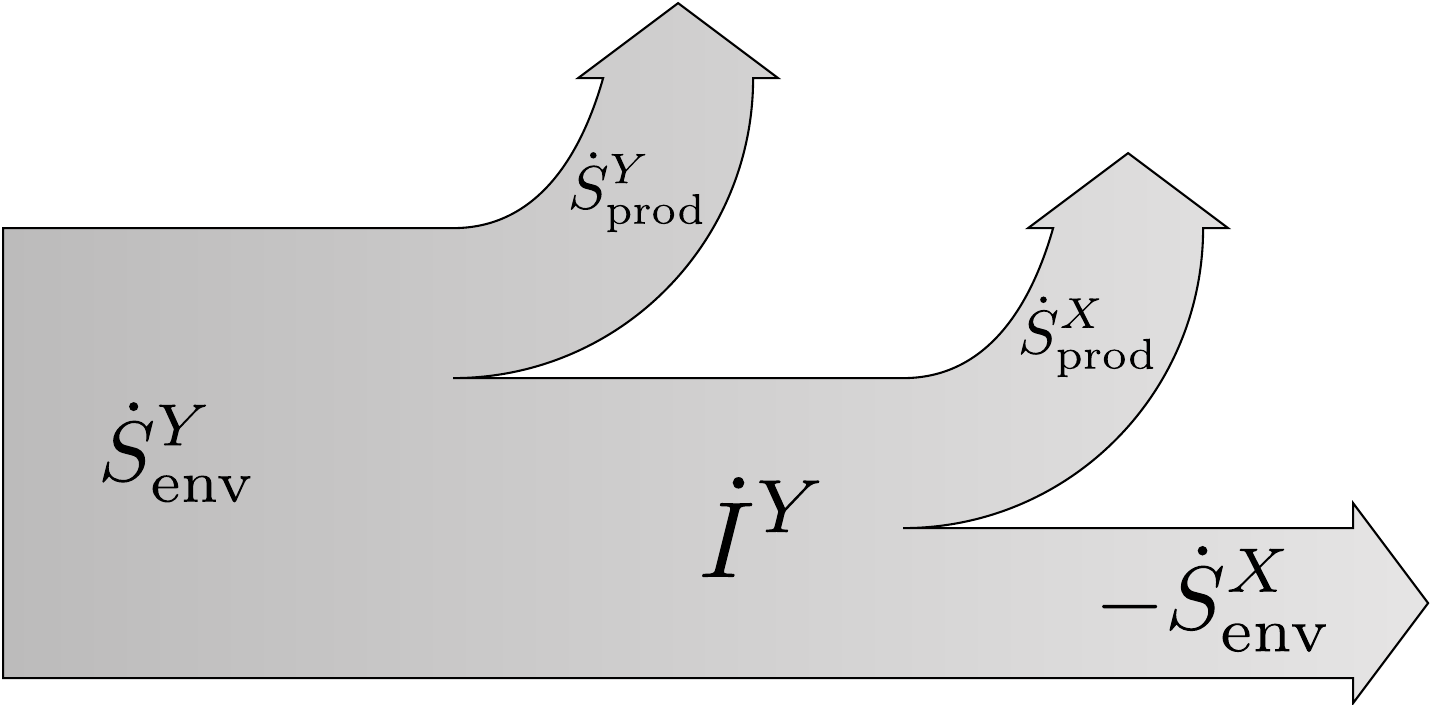}
    \caption{The entropic Sankey diagram of a bipartite system. The increase of entropy in the $Y$-environment, $\dot S^Y_{\rm env}>0$, is used to decrease the entropy in the $X$-environment, $\dot S^X_{\rm env}<0$. The transfer of this capacity to decrease entropy from system $Y$ to system $X$ occurs through the information flow $\dot I^Y>0$, while the entropy production in each system can be interpreted as leaks in this transfer \cite{Horowitz2014Jul}. All the quantities in the diagram are positive and represent the transfer of a conserved quantity. For instance, $\dot S^Y_{\rm env}=\dot S^Y_{\rm prod}+\dot I^Y$. Hence, the diagram is a graphic representation of the local second laws \eqref{secondflows} for the specific case $\dot S^Y_{\rm env}>0$, $\dot I^Y>0$ and $\dot S^X_{\rm env}<0$. }
    \label{fig:sankey}
\end{figure}

This use of thermodynamic resources can be represented by the flow diagram depicted in figure \ref{fig:sankey}, analogous to the so-called Sankey diagrams that represent the transfers and leaks of a given resource in engines and other processes and devices. In our case, the increase of entropy in the $Y$-environment is transferred to system $X$ through the information flow $\dot I^Y$ and used to decrease the entropy of the $X$-environment. The entropy production in each subsystem, $\dot S^X_{\rm prod}$ and $\dot S^Y_{\rm prod}$, can be considered leaks in this transfer of entropy.
However, this entropic Sankey diagram differs from the more usual one for engines, which describes transfers and losses of energy. In the entropic diagram of figure \ref{fig:sankey}, the resource is the increase of entropy in one of the reservoirs, $\dot S^Y_{\rm env}>0$. In a chemical motor, for instance, the reservoir could be a chemostat that indeed provides free energy to the motor; however, in a thermal motor, the reservoir where the entropy increases is the cold bath, where heat is dissipated. Moreover, the output in the diagram, $-\dot S_{\rm env}^X$ does not always coincide with or is proportional to the functional purpose of the machine, namely, the work in an engine or the extracted heat in a heat pump. An extreme case is the heat flow between two thermal baths in contact. This is not clearly a useful thermal machine; however, one could draw an entropic Sankey diagram where the resource is the entropy increase in the cold bath and the output is the entropy decrease in the hot bath. Despite this, the diagram is useful to analyze the performance of certain devices, as we show in the next section.

For isothermal chemical machines, the local second laws \eqref{secondflows} can be combined with the conservation of energy in the steady state to find a new interesting relation \cite{grelier2023,ehrich2023,Barato2017,large2021}. 
The change in the energy of the global bipartite system due to the evolution of system $Y$ is \footnote{In Refs. \cite{grelier2023,ehrich2023,large2021}, this quantity is denoted by $\dot W^{Y\to X}$ and interpreted as {\em transduced work} from system $Y$ to $X$.}
\begin{equation}
\dot E^Y\equiv \sum_{x,y<y'} J^x_{y\to y'}\left[E_{(x,y')}-E_{(x,y)}\right],
\end{equation}
where $E_{(x,y)}$ is the energy of state $(x,y)$.
As in \eqref{firstlaw}, this energy change  can be split into heat and chemical work:
\begin{equation}
\dot E^Y =\dot Q^Y+\dot W_{\rm chem}^Y,
\end{equation}
where the heat is related to the entropy change in the environment as  $T\dot S^Y_{\rm env}=-\dot Q^Y$. The same argument holds for the energy change due to the evolution of system $X$: $\dot E^X =\dot W_{\rm chem}^X-T\dot S^X_{\rm env}$. In the steady state, the average energy is constant, hence, $\dot E^X+\dot E^Y=0$. Then, one can express the entropy changes in terms of the chemical works and $\dot E^Y$:  $T\dot S^X_{\rm env}=W_{\rm chem}^X+\dot E^Y$  and $T\dot S^Y_{\rm env}=W_{\rm chem}^Y-\dot E^Y$. Introducing these expressions into the two inequalities \eqref{secondflows}, one obtains \cite{grelier2023,ehrich2023}
\begin{equation}\label{isomachines}
    \dot W_{\rm chem}^Y \geq \dot E^{Y} + T\dot I^Y \geq -\dot W^X_{\rm chem}.
\end{equation}

The local second laws \eqref{secondflows} and its particularization to isothermal machines \eqref{isomachines}  allow one to analyze separately the performance of each subsystem.
They have been applied to coupled quantum dots \cite{Horowitz2014Jul}, Brownian particles \cite{Allahverdyan2009Sep}, sensors \cite{hartich2016}, chemical networks \cite{penocchio2022,amano2022}, biological motors \cite{lathouwers2022,takaki2022}, and other stochastic systems \cite{grelier2023,ehrich2023}. Here we illustrate the theory with a novel application to nanomachines that couple electron transport and the oscillations of carbon nanotubes.

\section{A case study: nanotube oscillations induced by electron transport}
\label{sec:nanotubes}

In recent decades, several experimental realizations of nanoresonators activated by photons or electric transport have been implemented. These experimental setups are relevant in various applications including information processing \cite{LaHaye2009}, sensors \cite{Bachtold2013,Bachtold2018} and entanglement manipulation \cite{OConnell2010Apr, Lehnert2013,Lehnert2017}.
In the context of stochastic and quantum thermodynamics, nanoresonators have been used to devise nanomotors, where electron transport activates coherent oscillations in a carbon nanotube \cite{wen2020coherent}, as well as refrigerator engines \cite{urgell2020cooling}, where the electron transport damps the motion of the oscillator, consequently cooling the carbon nanotube. Both regimes are autonomous and have been analyzed using stochastic thermodynamics without involving the concept of information \cite{Nazarov2007,Wachtler2019Aug,wachtler2019stochastic}.

In most implementations, the nanoresonator's electrical conductivity depends on the position of the oscillator. Hence, the oscillation modulates the electron current, which at the same time drives the oscillator, 
creating a non-linear back action between electron transport and mechanics.

\begin{figure}
    \centering
    \includegraphics[width=.4\linewidth ]{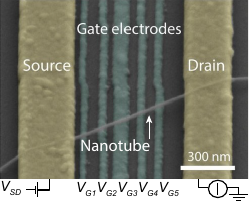}
    \qquad\quad
    \includegraphics[width=.4\linewidth ]{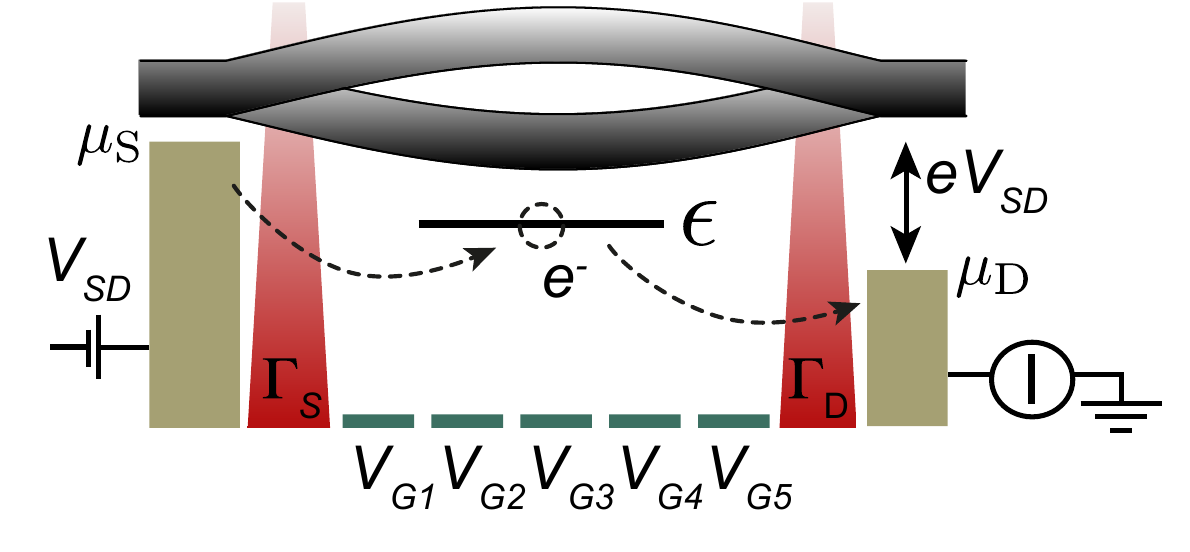}
    \caption{The carbon nanotube resonator. {\em Left:}  Micrograph of the device used in Ref.~\cite{Laird2018}. The nanotube  (thin straight line) lies over two gold electrodes, source and drain. Below the nanotube, five gate electrodes ($V_{G1}-V_{G5}$) create a confining electrostatic potential inducing a quantum dot within the nanotube. {\em Right:} Scheme of the device. The electrochemical potential of the quantum dot $\epsilon$ changes with the amplitude of the oscillation. The electrons' energies in the electrodes are distributed according to Fermi distributions with electrochemical potentials $\mu_{\rm S}$ (source) and $\mu_{\rm D}$ (drain), represented as gold rectangles. The voltage between the source-drain electrodes is $V_{\rm SD}= (\mu_{\rm S}-\mu_{\rm D})/e$, where $e$ is the elementary charge. Electrons tunnel between the dot and the source-drain electrodes through potential barriers. They are represented by the two red trapezoids, indicating that the corresponding tunneling rates, $\Gamma_{\rm S,D}(\epsilon)$, depend on $\epsilon$ and, in turn, on the position of the nanotube. }
    \label{fig:SER}
\end{figure}

Here, we study an experimental implementation based on fully suspended carbon nanotubes~\cite{Laird2012}. The left panel of figure~\ref{fig:SER} shows a micrograph of a typical device where a single carbon nanotube is suspended between two source-drain electrodes. Underneath the carbon nanotube, an array of five gate electrodes is used to control the transport properties of this device.
With an appropriate choice of gate voltages, the five gates electrodes create an electrostatic potential along the nanotube defining a quantum dot~\cite{ Nazarov2007,Ilani2014}. At sufficiently low temperatures, quantum dots display discrete energy transitions, and individual electron transport can be observed. The quantum dot population is thus a random variable that takes only two values, $p = 0,1$.  Applying a bias voltage between the source and drain electrodes, $eV_{\rm SD} = \mu_{\rm S} - \mu_{\rm D}$, where $e$ is the elementary charge, creates a measurable current that is proportional to the electron tunneling rates to the source and drain electrodes, $\Gamma_{S}$ and $\Gamma_{D}$, respectively.



\subsection{Dynamics}
\label{sec:dot_dynamics}

In this setup, mechanical oscillations are described by the amplitude $x$  of the fundamental vibration mode of the nanotube, and its velocity $v$. The evolution of the pair $(x(t),v(t))$ obeys the  underdamped Langevin equation \cite{Laird2019}
\begin{equation}
    \begin{split}
        m\dot v(t) &= -m\Omega^2 x(t) - m\frac{\Omega}{\mbox{\sf Q}}v(t) - gp(t) 
       + \xi(t),\\
        \dot x (t) &= v(t).
    \end{split}\label{eq:Langevin}
\end{equation}
Here $\Omega$ and $m$ are the frequency and mass of the vibration mode, respectively. The friction coefficient $\gamma\equiv m\Omega/\mbox{\sf Q}$ is given in terms of the quality factor $\mbox{\sf Q}$, which approximately equals the number of periods for an oscillation to decay due to friction. The term $-gp(t)$ represents the force experienced by the electric charge confined within the nanotube due to the gate electrostatic field,  $g$ being a coupling constant. The Gaussian white noise  $\xi(t)$ accounts for thermal fluctuations. Its average vanishes and the correlation reads $\langle\xi(t)\xi(t')\rangle = 2\gamma kT_{\rm osc}\delta(t-t')$, where $T_{\rm osc}$ is the temperature of the oscillator.

In turn, the electrochemical potential energy of the quantum dot $\epsilon$ is capacitively coupled to the nanotube position $x$, due to the electrostatic field created by the gates. To be consistent with the force $-gp(t)$ in \eqref{eq:Langevin}, this energy must read
\begin{equation}
    \epsilon(x) = \epsilon_0 + gx.
    \label{eq:capacitive_coupling}
\end{equation}

\begin{figure}
    \centering
    \includegraphics[width=.55\linewidth ]{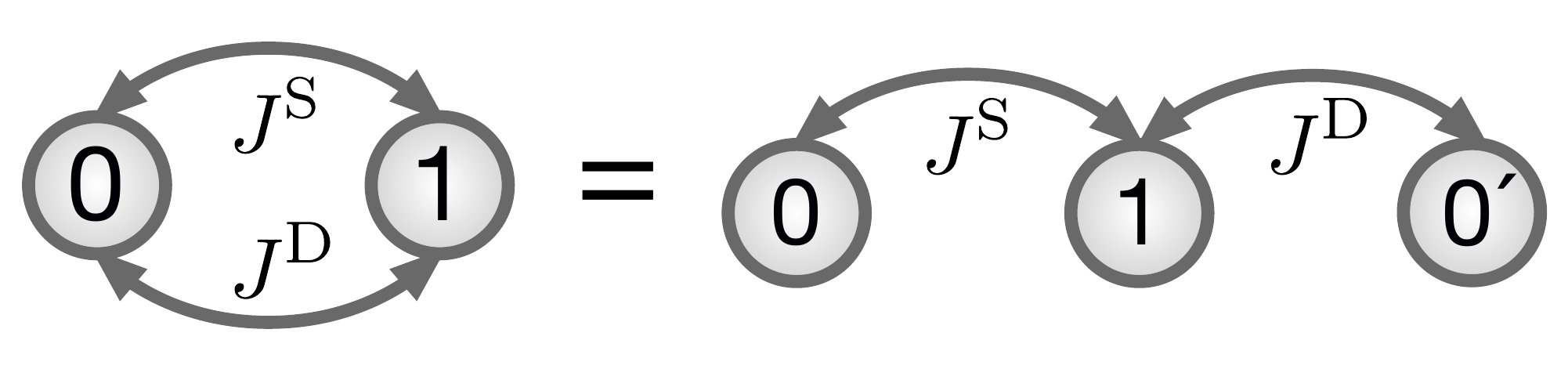}
    \caption{Markov chain describing the occupation of the quantum dot $p(t)$. One must distinguish between the transitions due to electron exchanges with the source and drain electrodes, respectively. Each transition obeys local detailed balance and induces the corresponding current $J^{\rm S,D}_{0\to 1}$. The resulting Markov chain is equivalent to a system with three states $0$, $1$, and $0'$, with periodic boundary conditions that identify the two states $0$ and $0'$.}
    \label{fig:two_states}
\end{figure}

The exchange of electrons between the dot and the electrodes is due to tunneling through potential barriers, as sketched in the right panel of figure \ref{fig:SER}. The barriers are characterized by tunneling rates $\Gamma_{\rm S,D}(\epsilon)$, which may depend on the dot energy $\epsilon$ \cite{meerwaldt2012}. The subindex denotes respectively the corresponding electrode, source (S) or drain (D). The occupation number $p(t)$ 
is a Markov chain with two states, $0$ or $1$, and the transitions between these two states are due to the exchange of electrons between the dot and the electrodes \cite{esposito2007}.  To preserve local detailed balance, it is necessary to distinguish between transitions due to the exchange of electrons with each electrode, as shown in figure \ref{fig:two_states}. The transition rates from $p=0$ to $p=1$ are $\Gamma^{\rm S,D}_{0\to 1}=\Gamma_{\rm S,D}(\epsilon)f_{\rm 
S,D}(\epsilon)$, where $f_{\rm S,D}(\epsilon)=1/\left[1+e^{\beta_{\rm dot}(\epsilon-
\mu_{\rm S,D})}\right]$ is the  Fermi function of each electrode at inverse temperature  $\beta_{\rm dot}=1/(kT_{\rm dot})$ and chemical potential $\mu_{\rm S,D}$. Notice that the temperature $T_{\rm dot}$ of the electrodes\footnote{We denote by $T_{\rm dot}$ the temperature of the electrodes to be consistent with our notation for entropy production in the environment of each subsystem: $T_{\rm dot}$ is the temperature of the environment of the dot, which consists of the two electrodes. The dot itself is in most applications out of equilibrium and its temperature (which could be defined through the ratio between the probability of being empty and occupied) has no clear physical meaning.} can be different from the oscillator temperature, $T_{\rm osc}$. On the other hand, if the dot is occupied, the electron leaves it at a rate $\Gamma^{\rm S,D}_{1\to 0}=\Gamma_{\rm S,D}(\epsilon)\left[1-f_{\rm 
S,D}(\epsilon)\right]$. These transition rates obey detailed balance:
\begin{equation}
\frac{\Gamma^{\rm S,D}_{0\to 1}}{\Gamma^{\rm S,D}_{1\to 0}}=
\frac{f_{\rm S,D}(\epsilon)}{1-f_{\rm S,D}(\epsilon)}=
e^{-\beta_{\rm dot} (\epsilon-\mu_{\rm S,D})}.
\end{equation}

The system exhibits a rich phenomenology, including self-oscillations sustained by electron currents \cite{Laird2019}, if the barriers and the tunneling rates  $\Gamma_{\rm S,D}(\epsilon)$ depend on the energy. In fact, in a realistic setup, the resulting barriers are always inhomogeneous because they are very sensitive to the geometry and electrostatic properties of the device, such as local potential disorder. In a typical dot configuration, they can be parametrized as \cite{meerwaldt2012}
\begin{equation}
    \Gamma_{\rm S,D}(\epsilon) = \gamma_{\rm S,D} e^{\alpha_{\rm S,D}\epsilon/g}
    \label{eq:barriers}
\end{equation}
where $\gamma_{\rm S,D}$ and $\alpha_{\rm S,D}$ are constant parameters of the device.

One can also define two probability currents, depending on $x$ and $v$:
\begin{equation}
\begin{split}
J^{{\rm S,D}}_{0\to 1}(x,v;t)=-J^{{\rm S,D}}_{1\to 0}(x,v;t)=\rho(x,v,0;t)\Gamma^{\rm S,D}_{0\to 1}(x)-
\rho(x,v,1;t)\Gamma^{\rm S,D}_{1\to 0}(x)
\end{split}
\end{equation}
where we have written explicitly the dependence of the transition rates on $x$ through the energy of the dot: 
$\epsilon(x)=\epsilon_0+gx$.

The dynamics of the whole system is governed by the Langevin equation \eqref{eq:Langevin} together with the stochastic jumps of the dot occupation $p(t)$. They induce an evolution equation for the probability distribution $\rho(x,v,p;t)$, which is a combination of a Fokker-Planck and a master equation:
\begin{equation} \label{fokker-planck}
        \frac{\partial\rho(x,v,p;t)}{\partial t} = -\frac{\partial v\rho}{\partial x}
        -\frac{\partial (F/m)\rho}{\partial v}
        +\frac{\Omega kT_{\rm osc}}{m\mbox{\sf Q}}\,\frac{\partial^2 \rho}{\partial v^2}+(2p-1)\left[J^{\rm S}_{0\to 1}+J^{\rm D}_{0\to 1}\right],
\end{equation}
where $F(x,v,p;t)=-m\Omega^2x-m\Omega v/\mbox{\sf Q}-gp
$ is the deterministic force acting on the nanotube.

In terms of probability currents, the equation can be written as
\begin{equation} \label{fokker-planckj}
        \frac{\partial\rho(x,v,p;t)}{\partial t} = -\Vec{\nabla} \cdot \left[\Vec{J}^{\rm rev}_p(x,v;t)+
        \Vec{J}^{\rm irrev}_p(x,v;t)\right]+(2p-1)\left[J^{\rm S}_{0\to 1}+J^{\rm D}_{0\to 1}\right],
\end{equation}
where the gradient is defined on the plane $(x,v)$, i.e., $\Vec{\nabla}\equiv(\partial_x,\partial_v)$.
In this expression, we have split the probability current associated with the nanotube motion into a conservative or reversible term and an irreversible contribution \cite{fischer2018}:
\begin{equation}
\begin{split}
    \Vec{J}^{\rm rev}_p(x,v;t)&\equiv \frac{1}{m}
    \left(\begin{array}{c} mv \\  -m\Omega^2x-gp
    \end{array}\right)\rho(x,v,p;t)
    \\
    \Vec{J}^{\rm irrev}_p(x,v;t)&\equiv -\frac{\Omega}{m\mbox{\sf Q}}
    \left(\begin{array}{c} 0 \\  \partial_v\left[mv^2/2+ kT_{\rm osc}\ln \rho \right]\end{array}\right)\rho(x,v,p;t) . 
\end{split}
\end{equation}

\subsection{Thermodynamics and information flows}

Here we proceed as in subsection \ref{sec:entropymaster}: we calculate the entropy production rate, $\dot S_{\rm prod}(t)$, as the sum of the change in the Shannon entropy of the system $\dot S(t)$ plus the change in the thermodynamic entropy of the environment $\dot S_{\rm env}(t)$. The contribution of the dot occupation to $\dot S_{\rm env}(t)$, i.e., the rate of change in the entropy of the electrodes,  can be expressed as in \eqref{senvxmark}:
\begin{align}
  \dot  S^{\rm dot}_{\rm env}(t)&=k\int dxdv \, \left[ J^{\rm S}_{0\to 1}\ln\frac{\Gamma^{\rm S}_{0\to 1}}{\Gamma^{\rm S}_{1\to 0}}+J^{\rm D}_{0\to 1}\ln\frac{\Gamma^{\rm D}_{0\to 1}}{\Gamma^{\rm D}_{1\to 0}}\right]\nonumber \\
&=\frac{1}{T_{\rm dot}}\int dxdv \, \left[ J^{\rm S}_{0\to 1}(x,v;t)(\mu_S-\epsilon(x))+
J^{\rm D}_{0\to 1}(x,v;t)(\mu_D-\epsilon(x))\right].\label{sdot}
\end{align}

To proceed, we have to quantify the change in the entropy of the environment due to mechanical oscillations. A way to do this is to calculate the variation of energy in the nanotube 
and identify the heat and work terms. Using the Fokker-Planck equation \eqref{fokker-planck} and integrating by parts, one can write the change of energy as
\begin{align}
  \dot E(t)&\equiv   \frac{d}{dt} \left\langle \frac{mv^2}{2}+\frac{m\Omega^2x^2}{2}+p\epsilon(x) \right\rangle\nonumber 
  \\
  &= \sum_{p=0,1}\int dxdv \left[\frac{mv^2}{2}+\frac{m\Omega^2x^2}{2}+p\epsilon(x) \right]\frac{\partial \rho(x,v,p;t)}{\partial t}\nonumber 
  \\
  &= \frac{\Omega}{\mbox{\sf Q}}\left[kT_{\rm osc}-m\langle v(t)^{2}\rangle \right]+\int dxdv\,\epsilon(x)\left[J^{{\rm S}}_{{0\to 1}}(x,v;t)
+J^{{\rm D}}_{{0\to 1}}(x.v;t)\right]
  \nonumber 
  \\
  &=
   \dot Q_{\rm osc}(t)+\dot Q_{\rm dot}(t)+\dot W_{\rm chem}(t)\label{Ebalance}
\end{align}
where 
the heat from the thermal bath of the oscillator is \cite{fischer2018,parrondo1996}
\begin{equation}
\dot Q_{\rm osc}(t)=-T_{\rm osc}\dot S_{\rm env}^{\rm osc}(t)=\frac{\Omega}{\mbox{\sf Q}}\left[kT_{\rm osc}-m\langle v(t)^{2}\rangle \right].\label{Qosc}
\end{equation}
As in \eqref{firstlaw}, the contribution of the electron current to the total energy change in \eqref{Ebalance}
is the sum of the heat transferred from the electrodes
$\dot Q_{{\rm dot}}(t)=
-T_{\rm dot}\dot S_{\rm env}^{{\rm dot}}(t)$, 
with $\dot S_{\rm env}^{\rm dot}(t)$ given by \eqref{sdot}, and the chemical work
\begin{equation}
\dot W_{\rm chem}(t)=\int dxdv\,\left[ J^{\rm S}_{0\to 1}(x,v;t)\mu_{\rm S}
+J^{\rm D}_{0\to 1}(x,v;t)\mu_{\rm D}\right].\label{Wchem}
\end{equation}

Now, we have all the elements to find the analogous to the local second laws \eqref{secondflows}. We first consider the total Shannon entropy of the system
\begin{align}
S(x,v,p;t)
&=-k\sum_{p=0,1}
 \int dxdv\, \rho(x,v,p;t)\ln\rho(x,v,p;t)\nonumber  \\
& =S_{\rm osc}(t)+S_{\rm dot}(t)- I(x,v;p)
\end{align}
where $S_{\rm osc}(t)$ and $S_{\rm dot}(t)$ are the Shannon entropies of each subsystem and the mutual information reads
\begin{equation}
I(x,v;p)=k\sum_{p=0,1}\int dxdv\, \rho(x,v,p;t)\ln\frac{\rho(x,v,p;t)}{\rho(x,v;t)\rho(p;t)}.
\end{equation}

The time derivative of the Shannon entropy can be split into a contribution from the dot and another from the oscillator
\begin{equation}
\dot S(x,v,p;t)=\dot S_{\rm osc}(t)+ \dot S_{\rm dot}(t)- \dot I^{\rm osc}(t)-\dot I^{\rm dot}(t)
\end{equation}
with
\begin{equation}
\dot I^{\rm osc}(t)\equiv -k\sum_{p=0,1}
 \int dxdv\, \Vec{\nabla} \cdot \left[\Vec{J}^{\rm rev}_p(x,v;t)+
        \Vec{J}^{\rm irrev}_p(x,v;t)\right]\ln\frac{\rho(x,v,p;t)}{\rho(x,v;t)}
\end{equation}
and
\begin{equation}
\dot I^{\rm dot}(t)\equiv k\sum_{p=0,1}
 \int dxdv\, (2p-1)\left[J^{\rm S}_{0\to 1}(x,v;t)+J^{\rm D}_{0\to 1}(x,v;t)\right]\ln\frac{\rho(x,v,p;t)}{\rho(p;t)}.
\end{equation}
Finally, we can sum up the Shannon entropy and the change in the entropy of the respective environments as in section \ref{sec:bipartite1}. After some cumbersome but straightforward algebra, we obtain the two local second laws:
\begin{equation}\label{seclawosc}
    \dot S_{\rm prod}^{\rm osc}(t)\equiv \dot S_{\rm osc}(t)- \dot I^{\rm osc}(t)+\dot S_{\rm env}^{\rm osc}(t)=\frac{\mbox{\sf Q}m}{\Omega T_{\rm osc}}\left\langle \left[ \frac{\Vec{J}^{\rm irrev}_p(x,v;t)}{\rho(x,v,p;t)}\right]^2\right\rangle \geq 0
\end{equation}
and
\begin{align}
\dot S_{\rm prod}^{\rm dot}(t)&\equiv \dot S_{\rm dot}(t)- \dot I^{\rm dot}(t)+\dot S_{\rm env}^{\rm dot}(t)
\nonumber \\ &=
k\int dxdv\, \sum_{\alpha={\rm S,D}}J^{\alpha}_{0\to 1}(x,v;t)\ln\frac{\rho(x,v,0;t)\Gamma^\alpha_{0\to 1}}{\rho(x,v,1;t)\Gamma^\alpha_{1\to 0}}\geq 0,\label{seclawdot}
\end{align}
which is a special case of \eqref{sprodX}. The sum of the two yields the second law for the global system:
\begin{equation}\label{seclawglobal}
    \dot S_{\rm prod}(t)=\dot S_{\rm osc}(t)+\dot S_{\rm dot}(t)-\dot I(t)-\frac{\dot Q_{\rm osc}(t)}{T_{\rm osc}}-\frac{\dot Q_{\rm dot}(t)}{T_{\rm dot}}\geq 0.
\end{equation}

Finally, we particularize these results when the device works in the stationary regime. The marginal entropies $S_{\rm osc}(t)$ and $S_{\rm dot}(t)$ are constant and the information flows obey $\dot I^{\rm dot}+\dot I^{\rm osc}=0$. Hence, the local second laws \eqref{seclawosc} and \eqref{seclawdot} read
\begin{equation}
    \begin{split}
        \dot S_{\rm prod}^{\rm osc}&=\dot S_{\rm env}^{\rm osc} - \dot I^{\rm osc} =-\frac{\dot Q_{\rm osc}}{T_{\rm osc}}-\dot I^{\rm osc}\geq 0\\
        \dot S_{\rm prod}^{\rm dot}& = \dot S_{\rm env}^{\rm dot} - \dot I^{\rm dot}=-\frac{\dot 
 Q_{\rm dot}}{T_{\rm dot}}+\dot I^{\rm osc}\geq 0.
    \end{split}
\end{equation}
If we use the conservation of energy, $\dot E(t)=0$ in \eqref{Ebalance}, $\dot Q_{\rm osc}+\dot Q_{\rm dot}+\dot W_{\rm chem}=0$, we obtain the following constraints for the performance of the device:
\begin{equation}\label{localsecondlawdevice}
    \begin{split}
        \dot S_{\rm prod}^{\rm osc}& = -\frac{\dot Q_{\rm osc}}{T_{\rm osc}}-\dot I^{\rm osc}\geq 0\\
        \dot S_{\rm prod}^{\rm dot}& = \frac{\dot 
 Q_{\rm osc}+\dot W_{\rm chem }}{T_{\rm dot}}+\dot I^{\rm osc}\geq 0.
    \end{split}
\end{equation}
In the stationary regime,
the chemical work  $\dot W_{\rm chem}$, defined in Eq.\eqref{Wchem}, adopts the familiar form:
\begin{equation}
    \dot W_{\rm chem} = eV_{\rm SD}J,
    \label{eq:Wchem_2}
\end{equation}
where $e$ is the elementary charge and
\begin{equation}
    J \equiv \int dx dv J^{\rm S}_{0\to 1}(x,v;t)=-\int dx dv J^{\rm D}_{0\to 1}(x,v;t)
\end{equation}
is the net electron current from the source to the drain electrode. Finally, the heat from the electrodes also simplifies in the stationary regime:
\begin{equation}
    \dot Q_{\rm dot}=-T_{\rm dot}\dot S_{\rm env}^{\rm dot}=-\dot W_{\rm chem}+ \int dxdv \, \left[ J^{\rm S}_{0\to 1}(x,v;t)+
J^{\rm D}_{0\to 1}(x,v;t)\right]\epsilon(x).\label{qdotNESS}
\end{equation}

The global second law is the sum of the inequalities \eqref{localsecondlawdevice}, which can be written as
\begin{equation}\label{secondlawpump}
    \dot W_{\rm chem}+\dot Q_{\rm osc} \left[1- \frac{T_{\rm dot}}{T_{\rm osc}} \right] \geq 0
\end{equation}
and establishes the main limitations of the device and its possible operating modes.

\begin{figure}
    \centering
    \includegraphics[width=.8\linewidth ]{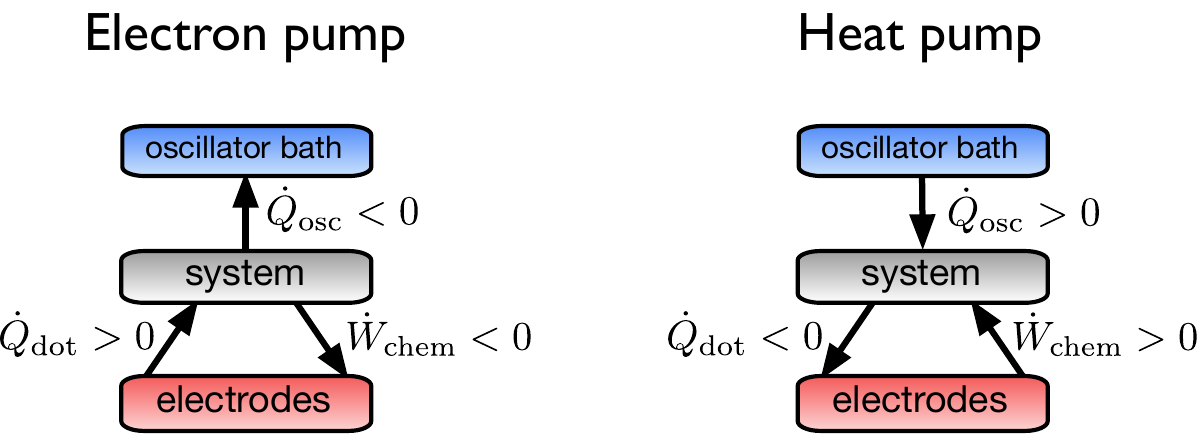}
    \caption{Energetics of the devices in the two regimes discussed in the text. The arrows indicate the direction of the energy transfer.  {\em Left:} the electron pump transfers electrons against the voltage performing work, $\dot W_{\rm chem}<0$. The energy is extracted from the electrodes at temperature $T_{\rm dot}$  ($\dot Q_{\rm dot}>0$) and part of this energy is dissipated to the bath of the oscillator ($\dot Q_{\rm osc}<0$) at lower temperature,  $T_{\rm osc}<T_{\rm dot}$,  to compensate the decrease of entropy in the electrodes. {\em Right:} a heat pump that transfers energy from the thermal bath of the oscillator ($\dot Q_{\rm osc}>0$) to the electrodes ($\dot Q_{\rm dot}<0$) against the temperature difference, i.e., when the oscillator bath is colder than the electrodes, $T_{\rm osc}<T_{\rm dot}$. To achieve this regime, the electron current must provide energy performing work, i.e. $\dot W_{\rm chem}>0$. 
    }
    \label{fig:2pump}
\end{figure}

In the next subsections, we study two modes of operation in the stationary regime, whose energetics are depicted in figure  \ref{fig:2pump}. In this figure, the arrows indicate the direction of the energy transfer. First, we analyze the device acting as an electron pump that moves electrons against the voltage: $\dot W_{\rm chem}<0$.  The energy to perform this work is extracted from a hot bath and part of it is dissipated as heat in a thermal bath, to comply with the second law of thermodynamics \eqref{secondlawpump}. More precisely, if  $\dot W_{\rm chem}<0$, then \eqref{secondlawpump} is satisfied only if $T_{\rm osc}< T_{\rm dot}$ and $\dot Q_{\rm osc}<0$ or if $T_{\rm osc}> T_{\rm dot}$ and $\dot Q_{\rm osc}>0$. Here we will only discuss the first case, where the entropy increases in the oscillator bath and decreases in the electrodes. The second mode of operation that we will consider below is the case where the electron current performs work, $\dot W_{\rm chem}>0$, and this work is used to pump heat from the thermal bath of the nanotube to the electrodes against the temperature gradient, i.e., $\dot Q_{\rm osc}>0$ and $\dot Q_{\rm dot}<0$ when $T_{\rm osc}\leq T_{\rm dot}$. Notice that the information flow is a necessary ingredient in these two regimes of operation.

\subsection{Electron pump}
\label{sec:autCNT}

\begin{figure}
    \centering
    \includegraphics[width=.9\linewidth ]{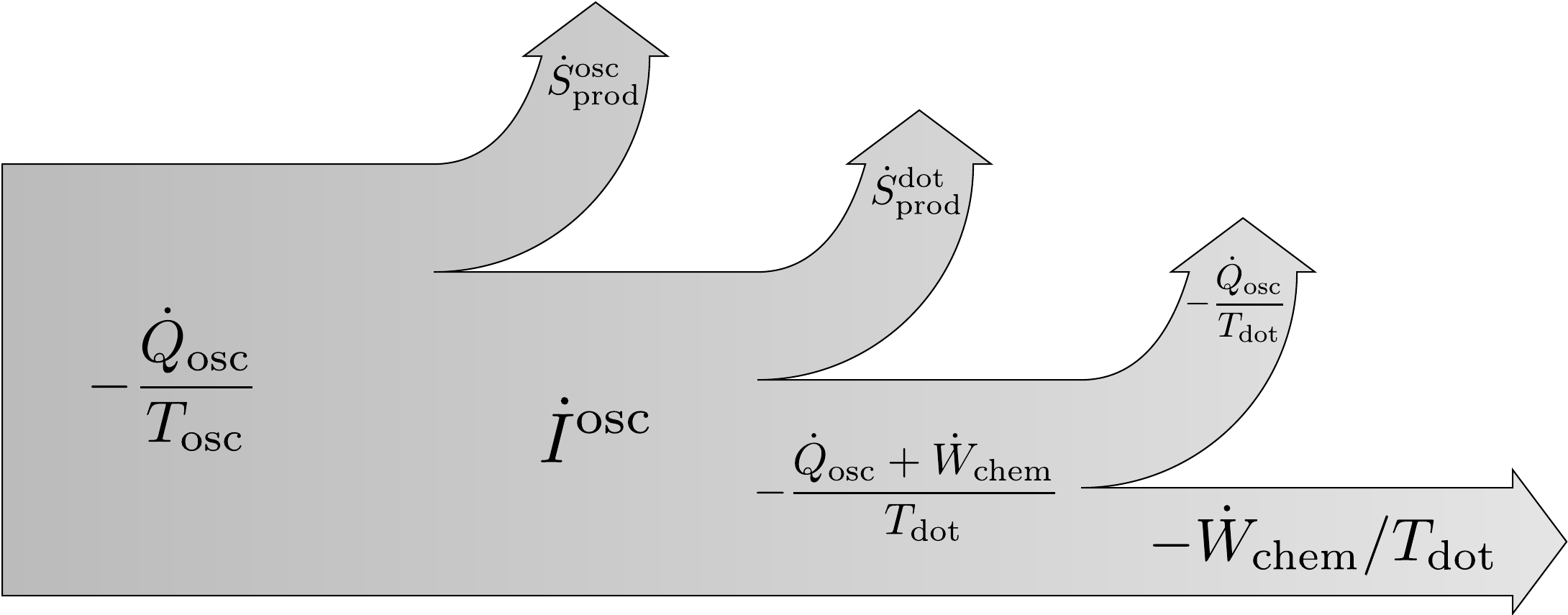}
    \caption{Sankey diagram of the electron pump. In this operation regime, heat is dissipated by the oscillator, $\dot Q_{\rm osc}<0$, and the dot pumps electrons against the voltage, $\dot W_{\rm chem}=eV_{\rm SD}J<0$. The energy comes from the heat transferred from the electrodes to the dot, $\dot Q_{\rm dot}=-\dot Q_{\rm osc}-\dot W_{\rm chem}>0$. To achieve this regime it is necessary that the environment of the oscillator is colder than the electrodes, $T_{\rm dot}>T_{\rm osc}$. 
    }
    \label{fig:sankeypump}
\end{figure}

In the electron pump regime, the device converts energy from the oscillations into a net current of electrons between the source and drain electrodes, even against an applied bias voltage. We further assume that  $T_{\rm osc}<T_{\rm dot}$. The energetics is depicted in the left panel of figure \ref{fig:2pump} and the corresponding Sankey diagram is shown in figure \ref{fig:sankeypump}, which is a special case of the diagram in figure \ref{fig:sankey}. Here, the nanotube plays the role of system $Y$ and the dot is system $X$. Hence, the information flow $\dot I^{\rm osc}=-\dot I^{\rm dot}$ is positive and the nanotube can be seen as a Maxwell demon operating on the quantum dot. In figure \ref{fig:sankeypump}, we have added a last stage to the diagram reflecting that the useful output is the chemical work.

From the global second law \eqref{secondlawpump}, it is straightforward to bound the efficiency of the pump:
\begin{equation}
   \eta\equiv \frac{-\dot W_{\rm chem}}{\dot Q_{\rm dot}}=\frac{\dot Q_{\rm dot}+\dot Q_{\rm osc}}{\dot Q_{\rm dot}}\leq  
\eta_{\rm Carnot}
\end{equation}
where $\eta_{\rm Carnot}\equiv 1- T_{\rm osc}/T_{\rm dot}$ is the Carnot efficiency. The equality is reached, as expected, when the total entropy production vanishes. Notice that the last ``leak'' in the Sankey diagram is not an entropy production but a loss due to the fact that not all the decrease in the entropy of the environment of the dot, $\dot Q_{\rm dot}/T_{\rm dot}=-(\dot Q_{\rm osc}+\dot W_{\rm chem})/T_{\rm dot}$, is used to perform chemical work.

\begin{figure}
    \centering
    \includegraphics[width=.6\linewidth ]{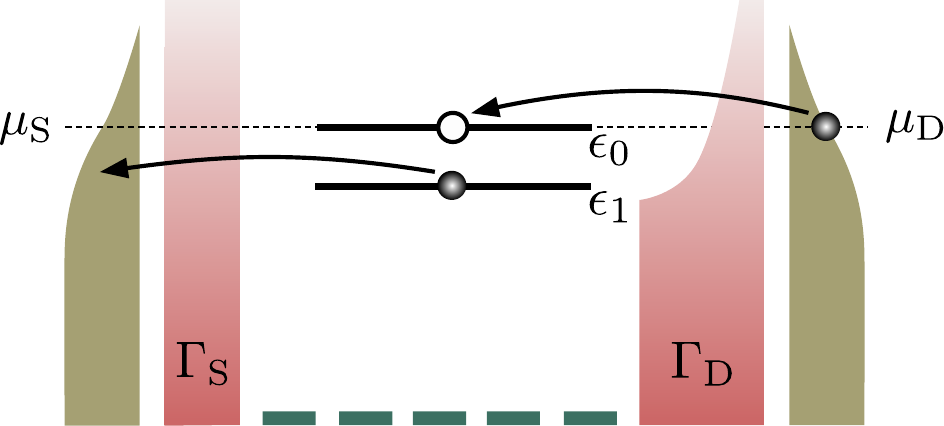}\qquad
    \includegraphics[width=.3\linewidth ]{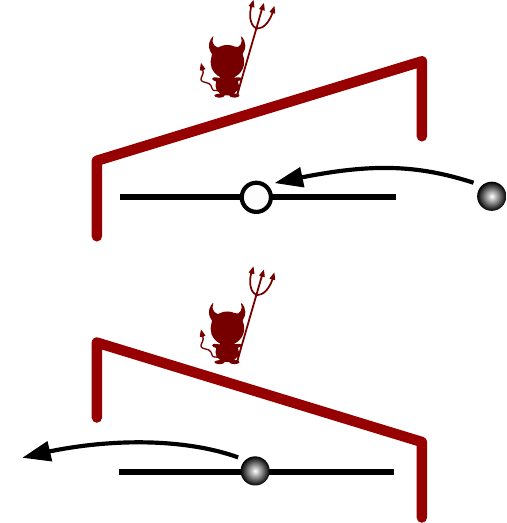}
    \caption{{\em Left:} Sketch of the electron pump mechanism at zero bias voltage, $V_{\rm SD}=0$ (i.e., for $\mu_{\rm S}=\mu_{\rm D}$). This mechanism is based on the inhomogeneity of the tunnel barriers $\Gamma_{\rm S,D}(\epsilon)$ and the shift of the equilibrium position of the oscillator when the dot is either filled ($p=1$) or empty ($p=0$). The electrochemical potential of the dot $\epsilon(x)=\epsilon_0+gx$ is tuned to match the Fermi levels of the source and drain electrodes when the oscillator is empty and in its equilibrium position $x_0=0$, i.e., $\epsilon(x_0)=\epsilon_0=\mu_{\rm S}=\mu_{\rm D}$. When the dot is filled with an electron ($p=1$), the equilibrium position of the oscillator is $x_1=-g/(m\Omega^2)$. At this position, the electrochemical potential of the dot is $\epsilon_1=\epsilon_0-g^2/(m\Omega^2)$. If $T_{\rm osc}$ is low enough and the tunnel barriers satisfy $\Gamma_{\rm S}(\epsilon_0)<\Gamma_{\rm D}(\epsilon_0)$ and $\Gamma_{\rm S}(\epsilon_1)>\Gamma_{\rm D}(\epsilon_1)$, then the device exhibits a net current from the drain to the source. 
    {\em Right:} The operation of the oscillator is sketched as a Maxwell demon that operates two trapdoors connecting the dot with the source and drain electrodes. The demon opens the right trapdoor and closes the left trapdoor when the dot is empty, whereas it opens the left trapdoor and closes the right one when the dot is filled. The operation generates a current from right to left.}
    \label{fig:electron_pump}
\end{figure}

As already mentioned, the information flow suggests that the oscllator acts as a demon to the dot. This interpretation can be made more precise by analyzing the dynamics of the device. Figure~\ref{fig:electron_pump} sketches the operation of the electron pump for the case of zero bias $V_{\rm SD}=0$, i.e. when the electrochemical potentials of the source and drain electrodes are the same, $\mu_{\rm S}=\mu_{\rm D}$. Recall that the amplitude of the nanotube behaves as a particle in a harmonic potential $m\Omega^2x^2/2+gpx$, which depends on whether the dot is filled ($p=1$) or empty ($p=0$). The equilibrium position of these two potentials is, respectively, $x_0=0$ and $x_1=-g/(m\Omega^2)$, for which the electrochemical potential of the dot $\epsilon(x)=\epsilon_0+gx$ is, respectively, $\epsilon(x_0)=\epsilon_0$ and $\epsilon(x_1)=\epsilon_0-g^2/(m\Omega^2)\equiv \epsilon_1$. The pump regime is achieved when these values are close to the electrochemical potential of the source and drain electrodes. The figure shows the case $\epsilon_0=\mu_{\rm S}=\mu_{\rm D}$ and $\epsilon_0-\epsilon_1=g^2/(m\Omega^2)\lesssim kT_{\rm dot}$. Another necessary condition for the pump to work is that at least one of the tunnel barriers depends on energy, $\Gamma_{\rm S,D}(\epsilon)$. In the figure, we have assumed that the left barrier (source) is homogeneous, $\alpha_{\rm S}=0$, and that the right barrier (drain) narrows when the energy of the dot increases, i.e. $\alpha_{\rm D}>0$ and $\Gamma_{\rm D}(\epsilon)$ is an increasing function of $\epsilon$, according to equation \eqref{eq:barriers}.

When the dot is empty, the oscillator's equilibrium position is $x_0$ and the electrochemical potential of the dot is $\epsilon_0$. Since $\Gamma_{\rm S}(\epsilon_0)<\Gamma_{\rm D}(\epsilon_0)$, the dot is filled by electrons coming, preferentially, from the drain electrode. Once the dot is filled, the oscillator moves to $x_1$ and shifts the electrochemical potential of the dot to $\epsilon_1$. In this case, $\Gamma_{\rm S}(\epsilon_0)>\Gamma_{\rm D}(\epsilon_0)$ and the electron leaves the dot preferentially from the source electrode. Hence, the device exhibits a net current of electrons from the drain to the source, and this current persists against moderate values of the bias voltage $V_{\rm SD}=\mu_{\rm S}-\mu_{\rm D}\geq 0$. Notice that the oscillator acts as a Maxwell demon: it ``measures" $p=0,1$, i.e. detects whether the dot is filled or not, and changes the electrochemical potential of the dot in a way that facilitates the transfer of electrons from the right electrode when the dot is empty and to the left electrode when the dot is filled. We illustrate this operation in the right panel of figure \ref{fig:electron_pump}, depicting the demon opening and closing trapdoors that connect the dot with the source-drain electrodes in a manner that induces a current from right to left.

\begin{figure}
    \centering
    \includegraphics[scale = .56]{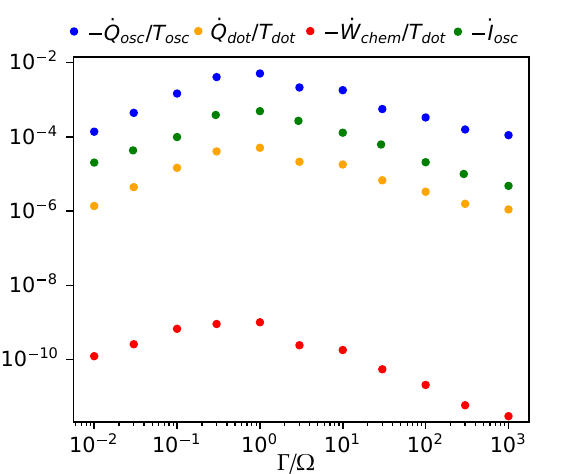}
    \includegraphics[scale = .56]{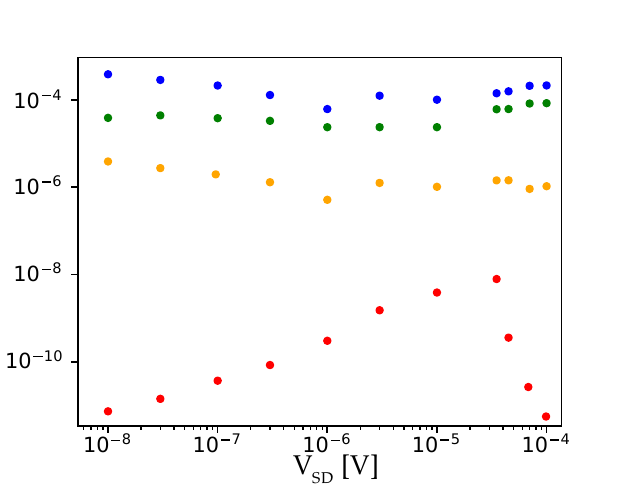}
    \caption{   The four entropy flows in the Sankey diagram of figure \ref{fig:sankeypump}: $-\dot Q_{\rm osc}/T_{\rm osc}$ (blue points), $\dot I^{\rm dot}$ (green),  $\dot Q_{\rm dot}/T_{\rm dot}$ (orange), and   $-\dot W_{\rm chem}/T_{\rm dot}$ (red), in units of Boltzmann constant per oscillation ($2\pi k/\Omega$). \textit{Top:} Flows as a function of the tunnel rates $\gamma_{\rm S} = \gamma_{\rm D} = \Gamma$, for $V_{\rm SD}=0.1\,\mu\mbox{V}$. {\em Bottom:} Flows as a function  of $V_{\rm SD}$ with $\Gamma/\Omega = 100$. In both panels $T_{\rm dot}/T_{\rm osc} = 10$, $\mu_S = \epsilon_0 = 0$, $kT_{\rm osc} = 0.01$~eV, $\Omega/2\pi=270$ MHz, $g=0.01$~eV/nm, $m =2\times 10^{-22}$ kg, $\mbox{\sf Q} = 1000$,  $\alpha_{\rm S}=0$, and $\alpha_{\rm D}=4~{\rm nm}^{-1}$.}
    \label{fig:electron_pump_flows}
\end{figure}

We have conducted numerical simulations of the device in the electron pump regime. In fig.\ref{fig:electron_pump_flows}, we show the four quantities appearing in the Sankey diagram of figure \ref{fig:sankeypump}: the heat flows $\dot Q_{\rm dot}$ and $\dot Q_{\rm osc}$, the chemical work $\dot W_{\rm chem}$, and the information flow, $\dot I^{\rm dot} = -\dot I^{\rm osc}$ for several values of $\Gamma\equiv\gamma_{\rm S,D}$ and bias voltage $V_{\rm SD}$.

The maximum power is reached when $\Gamma \simeq \Omega$, since in this case the demon operation, namely, the oscillations, and the electron transfer occur on similar time scales.
We observe that the chemical work is small compared with the heat dissipated, $\dot Q_{\rm dot}, \dot Q_{\rm osc} \gg \dot W_{\rm chem}$, that is, the electron pump mechanism is very inefficient from the point of view of thermodynamics, considering realistic device parameters. Notice however that the entropy production is relatively small compared to the last step in the Sankey diagram which is due to the different temperatures in the environments and limits the performance of the pump below the Carnot efficiency.

In the right panel of fig.\ref{fig:electron_pump_flows}, we show the entropic flows as a function of the bias voltage $V_{\rm SD}$. When the bias voltage is low, the electrons are pumped without resistance, performing no chemical work. If we increase the bias, the system starts performing work until $V_{\rm SD} \sim 10^{-5}$ V, where the bias voltage completely counteracts the electron pump, reversing the direction of the current. By converting $V_{\rm SD}$ to a temperature, we find that thermal and other sources of fluctuations must be reduced below approximately 110 mK in order for the pump current to be measurable in an experiment, a condition that is achieved in standard setups.



\subsection{Heat pump}

Our second example is a heat pump, whose energetics is represented in the right panel of figure \ref{fig:2pump}. In this regime, the direction of the electron current is the one given by $V_{\rm SD}$ and $T_{\rm osc}<T_{\rm dot}$. Hence, the work is positive, $\dot W_{\rm chem}>0$, and this energy is used to pump heat from the cold thermal bath affecting the oscillator ($\dot Q_{\rm osc}>0$) to the hot thermal bath, namely, the electrodes ($\dot Q_{\rm dot}<0$). This behavior can be seen as an {\em anti-Joule} effect since the current cools down the nanotube using its motion.

\begin{figure}
    \centering
    \includegraphics[width=0.65\linewidth ]{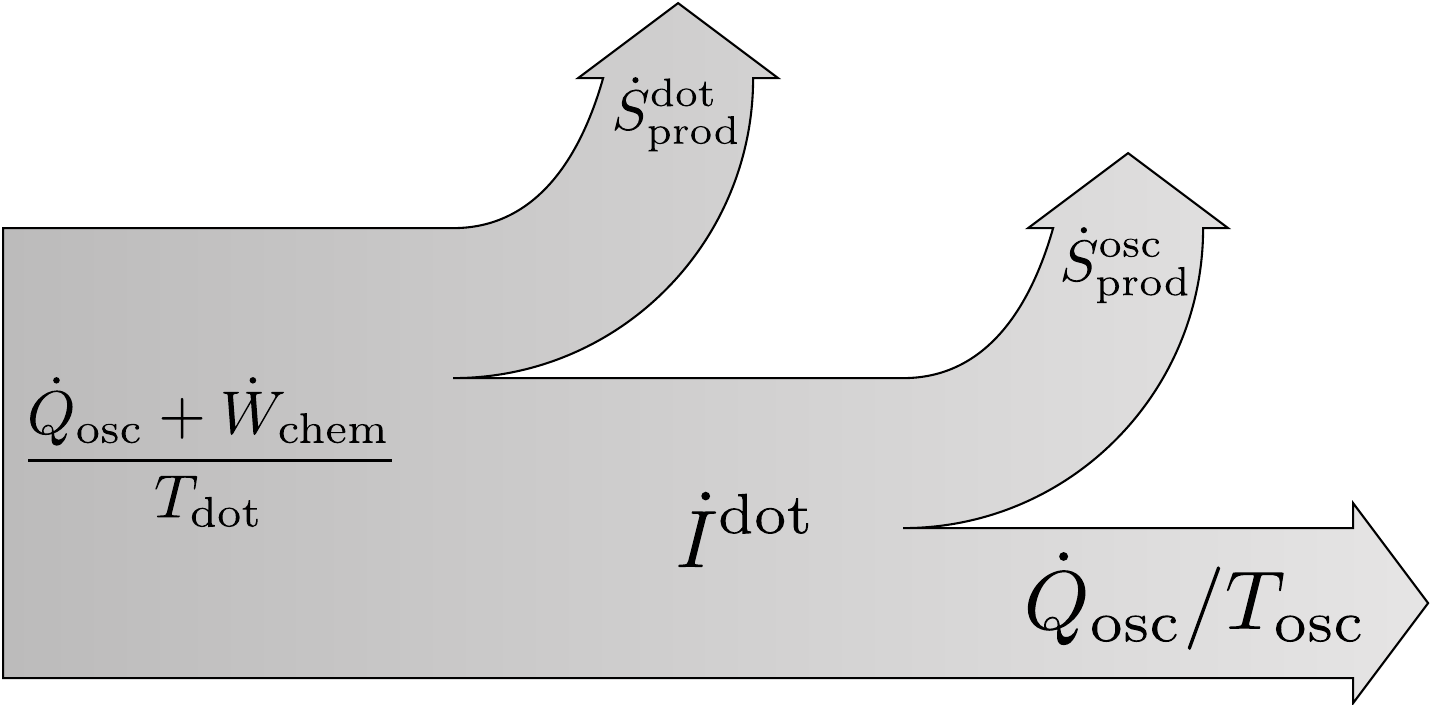}
    \caption{Sankey diagram of a heat pump. In this case, $\dot Q_{\rm osc}>0$ is the heat extracted from the cold bath, $T_{\rm osc}<T_{\rm dot}$, and the electron current performs a work $\dot W_{\rm chem}=eV_{\rm SD}J>0$}
    \label{fig:sankeyrefrigerator}
\end{figure}

The corresponding entropic Sankey diagram is depicted in figure \ref{fig:sankeyrefrigerator}. As in the case of the electron pump, the global second law imposes an upper bound to the efficiency of the machine \cite{adkins1983}:
\begin{equation}\label{secondlawheatpump}
    \eta\equiv \frac{\dot Q_{\rm osc}}{\dot W_{\rm chem}}\leq \frac{T_{\rm osc}}{T_{\rm osc}-T_{\rm dot}}
\end{equation}
which diverges when the two temperatures are equal. The refrigeration power is also bound by
\begin{equation}\label{qboundheatpump}
    \dot Q_{\rm osc}<\frac{kT_{\rm osc}\Omega}{\sf Q},
\end{equation} 
as can be deduced from \eqref{Qosc}. 
This inequality is more general and is verified by systems in which energy is extracted from a bath by means of a mechanical degree of freedom. The inequality is due to the fact that energy is extracted when the velocity fluctuations $\langle v^2\rangle$ are smaller than the fluctuations at equilibrium  $kT/m$  \cite{parrondo1996}.

\begin{figure}
    \centering
    \includegraphics[width=0.45\linewidth ]{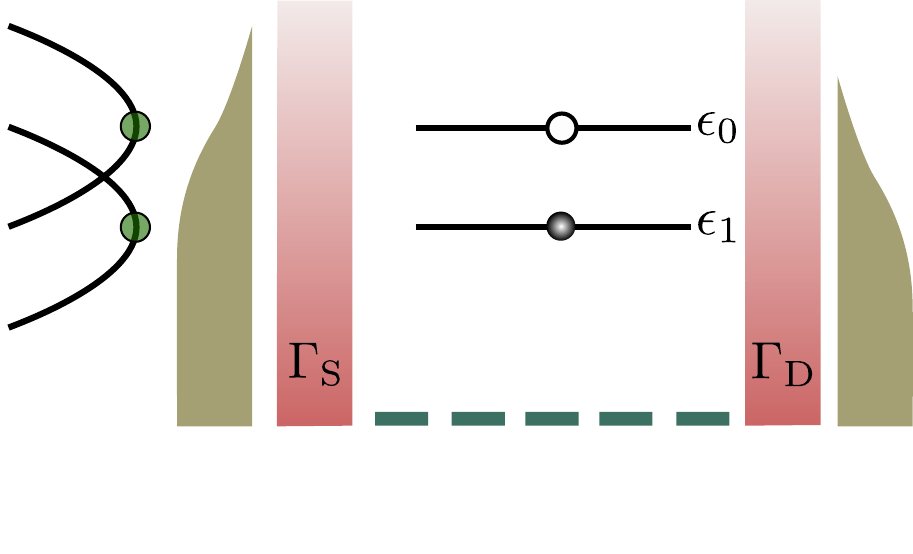}\qquad 
    \includegraphics[width=0.45\linewidth ]{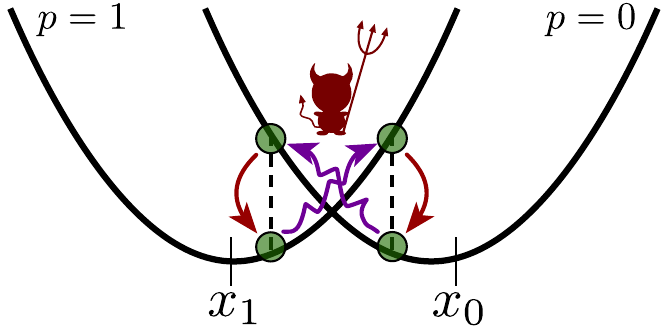}
    \caption{Sketch of the mechanism of the heat pump. {\em Left:} When the dot is empty, the oscillator fluctuates around $x_0$. However, a significant thermal fluctuation can push the oscillator downward, thereby increasing the likelihood of the dot being filled. When this happens,  the potential that the oscillator feels is shifted downward, and the oscillator transfers energy to the dot. A similar effect happens when a fluctuation drives the oscillator upwards and the dot loses its electron. {\em Right:} A Maxwell demon extracting energy from the environment of an oscillator acts in a similar way. When a thermal fluctuation drives the oscillator (green circle) far from the corresponding equilibrium position, i.e., to a point with high potential energy,  the demon shifts the potential reducing the energy of the oscillator. The demon is extracting the energy that the environment transfers to the oscillator through the fluctuations. The purple zigzag arrows indicate the motion of the oscillator due to fluctuations, and the red arrows indicate the change in 
    energy due to the switch in the potential carried out by the demon.}
    \label{fig:demonrefrigerator}
\end{figure}

The positive information flow now corresponds to the quantum dot, $\dot I^{\rm dot}=-\dot I^{\rm osc}>0$, indicating that in this regime, the dot functions as a Maxwell demon operating on the oscillator. In figure \ref{fig:demonrefrigerator}, we present an interpretation of the device's operation. 
In this case, it is more illustrative to consider how energy can be extracted from the thermal bath of an oscillator through feedback. The procedure is illustrated in the right panel of figure \ref{fig:demonrefrigerator}. 
In the device, when the quantum dot is empty ($p=0$), the oscillator moves around the equilibrium position $x_0$ thereby setting the electrochemical potential of the dot to $\epsilon_0$. In this configuration, if a fluctuation pushes the oscillator downward, the probability of the dot being filled increases. Consequently, for sufficiently large fluctuations, the dot is filled with an electron when the oscillator moves downward. This corresponds to the demon switching the potential to $p=1$, which changes the equilibrium position of the oscillator to $x_1$ and sets the electrochemical potential of the dot to $\epsilon_1$.
In figure~\ref{fig:demonrefrigerator}, fluctuations are depicted as purple zigzag arrows and the potential switch induced by the demon is represented by red arrows. The switch reduces the energy of the system, which is absorbed by the demon. Once the potential corresponds to $p=1$, the demon repeats the process, waiting now for a sufficiently large fluctuation to push the oscillator upwards closer to $x_0$. 



\begin{figure}
    \centering
    \includegraphics[scale = .55]{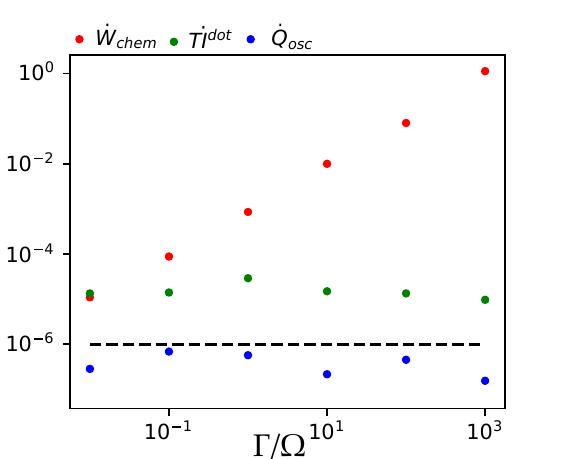}\quad
    \includegraphics[scale = .55]{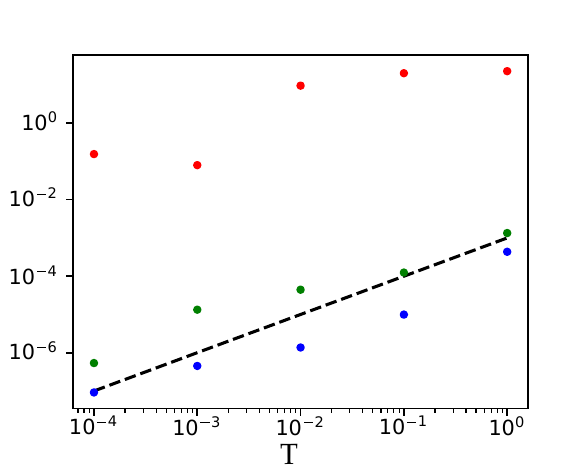}
    \caption{Heat flow $\dot Q_{\rm osc}$ (blue points), chemical work $\dot W_{\rm chem}$ (red points), and information flow $T_{\rm dot}\dot I^{\rm dot}$ (green points) in eV per oscillation: \textit{Left:} for several tunnel rates $\gamma_{\rm S} = \gamma_{\rm D} = \Gamma$ with $kT_{\rm osc} = kT_{\rm dot} = 0.001$~eV; \textit{Right:} for several temperatures $T_{\rm osc} = T_{\rm dot} = T$ with $\Gamma = 100\Omega$. The black dashed line is the bound \eqref{qboundheatpump}, i.e., $\Omega/\mbox{\sf Q}kT_{\rm  osc}$. The bias voltage is $V{\rm SD} = -2~{\rm mV}$ and the rest of the parameters are the same as in Fig.~\ref{fig:electron_pump_flows}.}
    \label{fig:cooling2}
\end{figure}

We have performed numerical simulations of the model for equal temperatures $T_{\rm osc}=T_{\rm dot}=T$. In this case, the second law \eqref{secondlawheatpump} does not provide an upper finite bound on the efficiency of the device, because a heat flow can be established with arbitrarily small work. However,  the refrigeration power $\dot Q_{\rm osc}$ is limited by equation \eqref{qboundheatpump}.
In fig.\ref{fig:cooling2} we depict the energy transfers: $\dot W_{\rm chem}$ (red), $T\dot I^{\rm dot}$ (green), $\dot Q_{\rm osc}$ (blue), and the bound $kT\Omega/{\sf Q}$ (dashed line) as a function of the tunneling rates $\Gamma$ and the temperature $T$. We observe that the highest entropy production occurs in the electron current, except in cases of low tunneling rates $\Gamma$.

\section{Conclusions}

We have provided a self-contained review of the definition and use of information flows in bipartite systems that exchange energy and particles with reservoirs. The main result is the two local second laws \eqref{secondflows}, which impose an extra constraint on the energetics of a bipartite system in a non-equilibrium steady state (NESS). As illustrated by the Sankey diagram in figure \ref{fig:sankey}, the information flow $\dot I^Y=-\dot I^X$ quantifies the transfer of negative entropy between the two subsystems, $X$ and $Y$. Moreover, it enables us to interpret the overall system as an information machine, where one of the subsystems plays the role of a Maxwell demon, measuring and operating on the other subsystem \cite{Horowitz2014Jul,hartich2014}. This interpretation holds the potential to enhance the design of nanomachines and the understanding of biological motors.

We have illustrated these ideas by analyzing a bipartite system consisting of a quantum dot in a carbon nanotube, where the mechanical oscillations of the nanotube are coupled to electron transport. Information flows allow one to interpret the device as an information machine. This interpretation provides hints for improving the efficiency or even achieving the desired functionality. In the case of the electron pump, for instance, the inhomogeneity of the tunnel barriers is a necessary condition for pumping electrons against the bias voltage. Mutual information reveals that this inhomogeneity is necessary to create a correlation between the population of the dot and the position of the nanotube.


\acknowledgement
We thank Jordan Horowitz and Jannik Ehrich for their thoughtful comments.
This research was supported by grant number FQXi-IAF19-01 from the Foundational Questions Institute Fund, a donor-advised fund of Silicon Valley Community Foundation. JMRP and JT-B  acknowledge financial support from Spanish Government through Grant FLUID (PID2020-113455GB-I00).
NA and FF acknowledge the support from the Royal Society (URF-R1-191150), from the European Research Council (ERC) under the European Union's Horizon 2020 research and innovation programme (grant agreement number 948932), and the European Commission via the Horizon Europe project ASPECTS (Grant Agreement No. 101080167).

\bibliographystyle{unsrt}
\bibliography{biblio.bib}


\end{document}